\documentclass[nofootinbib,preprintnumbers,superscriptaddress,amssymb]{revtex4}
\pdfoutput=1
\usepackage{graphicx}
\usepackage{slashed}
\usepackage{amsfonts}
\usepackage{dsfont}
\usepackage{enumerate}
\usepackage{color}
\usepackage{appendix}
\usepackage{amsmath}
\usepackage{bm}
\usepackage[paperwidth=210mm,paperheight=280mm,centering,hmargin=2.2cm,vmargin=2.2cm]{geometry}

\newcommand{\bea}{\begin{eqnarray}}
\newcommand{\eea}{\end{eqnarray}}

\def\X5sp{{\rm X}_5}
\def\Y3sp{{\rm Y}_3}
\def\Z3sp{{\rm Z}_3}

\def\be{\begin{equation}}
\def\ee{\end{equation}}
\def\bea{\begin{eqnarray}}
\def\eea{\end{eqnarray}}

\def\super3{{}^{(3)}}

\begin{document}

\title{Curvature perturbation in multi-field inflation \\with non-minimal coupling}

\author{Jonathan White}
\email{jwhite@yukawa.kyoto-u.ac.jp}

\author{Masato Minamitsuji}
\email{masato@yukawa.kyoto-u.ac.jp}

\author{Misao Sasaki}
\email{misao@yukawa.kyoto-u.ac.jp}
\affiliation{Yukawa Institute for Theoretical Physics,\\Kyoto University, Kitashirakawa Oiwakecho, Sakyo-ku, Kyoto 606-8502, Japan}

\begin{abstract}
In this paper we discuss a multi-field model of inflation in which generally all fields are non-minimally coupled to the Ricci scalar and have non-canonical kinetic terms.  The background evolution and first-order perturbations for the model are evaluated in both the Jordan and Einstein frames, and the respective curvature perturbations compared.  We confirm that they are indeed not the same - unlike in the single-field case - and also that the difference is a direct consequence of the isocurvature perturbations inherent to multi-field models.  This result leads us to conclude that the notion of adiabaticity is not invariant under conformal transformations.  
Using a two-field example we show that even if in one frame the evolution is adiabatic, meaning that the curvature perturbation is conserved on super-horizon scales, in general in the other frame isocurvature perturbations continue to source the curvature perturbation. We also find that it is possible to realise a particular model in which curvature perturbations in both frames are conserved but with each being of different magnitude.  These examples highlight that the curvature perturbation itself, despite being gauge-invariant, does not correspond directly to an observable.  The non-equivalence of the two curvature perturbations would also be important when considering the addition of Standard Model matter into the system.  
\end{abstract}

\preprint{YITP-12-37}

\maketitle

\section{Introduction}
An epoch of inflation in the early Universe has become widely accepted as one of the key ingredients in the standard model of cosmology \cite{Inflation}.  A specific model for inflation, however, is still yet to be determined, and current observational constraints can be satisfied by many of the models that have been proposed.  In spite of the agreement between current observations and the simplest of single-field inflation models, in the context of unifying or higher-dimensional theories, it is natural to consider a wider range of possibilities.  This might include making modifications to the kinetic term of the scalar field, such as in k-inflation \cite{k-inflation}, modifying the gravitational sector of the theory, such as in $f(R)$-gravity \cite{Felice}, introducing additional coupling between the gravity and matter sectors, such as in scalar-tensor theories of gravity \cite{Maeda} or introducing multiple scalar-fields \cite{WandsRev}.  Many of the proposed models do predict unique observable signatures in the statistics of the primordial perturbations they would produce, and it is therefore hoped that as more precise data becomes available we should be able to start constraining our model of inflation further.

The particular type of model that we consider here takes an action of the form 
\begin{equation}\label{JAc}
S = \int d^4x\sqrt{-g}\left\{f({\bm \phi})R - \frac{1}{2}G_{IJ}({\bm \phi})g^{\mu\nu}\partial_\mu\phi^I\partial_\nu\phi^J - V({\bm \phi})\right\},
\end{equation}
where $I,~J=1,...,N$ label $N$ scalar fields, potentially all fields are non-minimally coupled to gravity through the function $f({\bm \phi})$ - the vector argument ${\bm \phi}$ indicating the dependence of $f$ on all of the fields - and $G_{IJ}({\bm \phi})$ gives a non-canonical kinetic term.  $g^{\mu\nu}$, $g$ and $R$ are the 4-dimensional metric, its determinant and associated Ricci scalar, respectively, and $V(\bm{\phi})$ is some general potential.  We note here that $G_{IJ}({\bm \phi})$ can be interpreted as inducing a field manifold for which it is the metric.  This form of action is well motivated in the context of unifying theories, as the non-minimal coupling and non-canonical kinetic terms in \eqref{JAc} are also generic features of the 4-dimensional effective actions one obtains from higher-dimensional theories by way of compactification \cite{Copeland}.  

In trying to determine the primordial perturbations generated by the model \eqref{JAc}, the presence of non-minimal coupling between gravity and the multiple fields makes calculations rather more involved than the minimally coupled case.  In order to alleviate this problem, it is common practise to first make the conformal transformation 
\begin{equation}\label{cT}
g_{\mu\nu} = \Omega\tilde{g}_{\mu\nu},
\end{equation}          
where $\Omega>0$ to preserve causal structure.  With an appropriate choice of $\Omega$ we are able to recover the canonical Einstein-Hilbert form for the gravity part of the action, namely 
\begin{equation}\label{EAc}
S = \int d^4x\sqrt{-\tilde{g}}\left\{\tilde{R} -\frac{1}{2}S_{IJ}(\bm{\phi})\tilde{g}^{\mu\nu}\partial_\mu\phi^I\partial_\nu\phi^J - \tilde{V}(\bm{\phi})\right\},
\end{equation}     
where all quantities associated with this new metric carry a tilde and the quantities $S_{IJ}(\bm{\phi})$ and $\tilde{V}(\bm{\phi})$ will be given in Sec.\ref{efAnal}.  As such we have reduced the problem to a much more familiar one.  Given the standard form of the gravity sector in \eqref{EAc}, this choice of conformal frame is referred to as the Einstein frame.  If we consider introducing additional matter into the action \eqref{JAc} that is minimally coupled to $g^{\mu\nu}$, then this original frame is referred to as the Jordan frame, and test particles will follow geodesics of $g^{\mu\nu}$.  In the Einstein frame, however, this additional matter will not be minimally-coupled with $\tilde{g}^{\mu\nu}$, meaning that test particles will not follow geodesics of $\tilde{g}^{\mu\nu}$.

In making the conformal transformation \eqref{cT} all we have done is to re-label the metric.  Mathematically, therefore, at the classical level we are free to perform calculations in either frame, and any observable predictions should be the same \cite{ND+MS example}.\footnote{Note the importance of keeping track of the non-minimal coupling induced between matter and the scalar fields $\phi^I$ in the Einstein frame as a result of the conformal transformation.  It is this non-minimal coupling that leads to properties such as the space-time dependence of particle masses, which in turn leads to very different physical interpretations in the two frames.}  The physical interpretation in each frame, however, may be very different, and the debate as to which frame is ``the physical one" is a longstanding one.  As such, we must be very careful when attaching any physical meaning to quantities we calculate in one frame or the other that aren't directly observable.   

\begin{figure}[h!]
\centering
\includegraphics[width=0.6\textwidth]{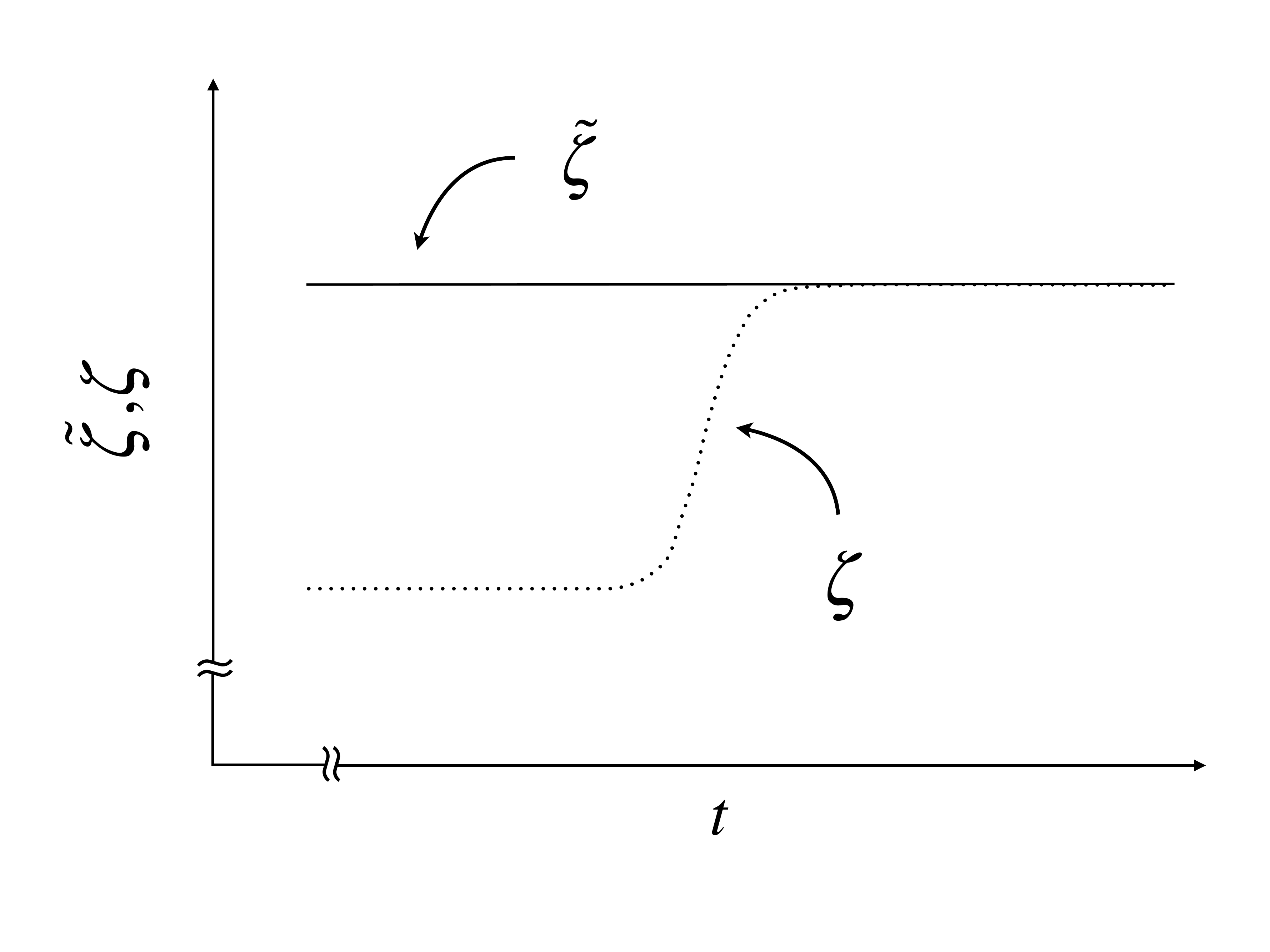}
\caption{A possible example of non-equivalent evolutions of the curvature perturbation in the Jordan (dotted curve, $\zeta$) and Einstein (solid line, $\tilde{\zeta}$) frames is shown, which will be considered in Sec.\ref{2fe}. In this example, in the initial stage curvature perturbations in both frames are conserved but are of different magnitude. Then, in the Jordan frame entropy perturbations start to source the curvature perturbation, whilst there is still no such sourcing in the Einstein frame. The background evolution in both frames eventually converges into a single adiabatic trajectory, along which $\zeta$ coincides with $\tilde{\zeta}$.}
\label{example}
\end{figure}

One quantity that we are particularly interested in calculating is the curvature perturbation on hypersurfaces of constant energy density, $\zeta$, as it is a gauge-invariant measure of the primordial perturbations that give rise to temperature fluctuations in the cosmic microwave background (CMB) and act as the seeds for structure formation.  Under rather general conditions in single-field models of inflation it is known that $\zeta$ is conserved on super-horizon scales \cite{Lyth}\cite{Naruko}.  It is also known in the single field case that $\zeta = \tilde{\zeta}$ not only to linear \cite{Makino}\cite{Tsujikawa} and second order \cite{Koh}, but to all orders in perturbation theory on super-horizon scales \cite{Gong}\cite{Chiba1}.  This means that we can freely perform our calculations in the Einstein frame, without worrying about how the final quantity should be related to the equivalent one in the original Jordan frame \cite{QiuYang}\cite{Kubota}.   In multi-field models, however, this is generally no longer the case, with both $\dot{\zeta}\neq 0$ and $\zeta \neq \tilde{\zeta}$.  The non-conservation of $\zeta$ on super-horizon scales is sourced by entropy perturbations as \cite{bs}
\begin{equation}
\dot{\zeta} = -\frac{H}{\rho+p}\delta p_{\rm{nad}},
\end{equation}
where $H$, $\rho$ and $p$ are the Hubble rate, background density and background pressure respectively, and $\delta p _{\rm{nad}}$ is the non-adiabatic pressure perturbation defined as 
\begin{equation}\label{nadPdef}
\delta p_{\rm{nad}}=\delta p - \frac{\dot{p}}{\dot{\rho}}\delta \rho.
\end{equation}
With $\zeta \neq \tilde{\zeta}$, this means that in general their evolutions will also differ, so that the idea of non-adiabaticity may be frame dependent.  In particular, it may be the case that whilst the curvature perturbation is conserved in one frame it is not in the other, i.e. $\delta \tilde{p}_{\rm{nad}} = 0\nLeftrightarrow \delta p_{\rm{nad}} = 0$.  This highlights the fact that attaching any physical meaning to the quantity $\zeta$ itself is somewhat arbitrary, as, despite being a gauge-invariant quantity, it is not directly observable.  Another possibility might be a situation where the curvature perturbation is conserved in both frames, but with each being of different magnitude.  During such a phase one might naively take the model to be effectively single-field, which would in turn imply the conformal equivalence of the curvature perturbation.  However, the presence of the isocurvature fields and non-minimal coupling means that in fact this is not necessarily the case.  If one assumes that an effectively single-field, adiabatic limit is reached before the present time, which means that eventually we do have $\zeta = \tilde{\zeta}$, then this leads to the possibility sketched in Fig. \ref{example}.   
Here, in the Einstein frame the curvature perturbation is conserved throughout the evolution, whilst in the Jordan frame one has a phase in which entropy perturbations source $\zeta$.  As such, we see that despite the final agreement of the two frames regarding the curvature perturbation, the interpretation of the evolution is very different.  

In this paper we explicitly calculate the curvature perturbation resulting from \eqref{JAc} in both the Jordan and Einstein frames and compare the final results.  In doing so we adopt the notion of a field manifold being induced by $G_{IJ}(\bm{\phi})$ in the Jordan frame and give the corresponding background and perturbed equations of motion for the fields in a covariant form \cite{Groot}\cite{Langlois}\cite{Tegmark}.  In the Einstein frame, as can be seen from \eqref{EAc}, we also find that we can naturally define a new metric, $S_{IJ}(\bm{\phi})$, that can again be used to express the equations of motion in a covariant form.  In the language of this geometric approach, one finds that in the Einstein frame $\dot{\tilde{\zeta}}=0$ when the background trajectory is a geodesic of $S_{IJ}(\bm{\phi})$ or when an adiabatic limit is reached, i.e. when the model becomes effectively single-field.  In the Jordan frame, however, such a geometric interpretation does not seem possible, and the only obvious situation where $\dot{\zeta} = 0$ appears to be when an effectively single-field adiabatic limit is reached.  In order to try and demonstrate the possibility of a scenario such as that outlined above, we then look to a two-field example in which $\zeta\neq\tilde{\zeta}$.  Given that we have a greater understanding as to when the curvature perturbation is conserved in the Einstein frame, we first impose $\delta \tilde{p}_{\rm{nad}}=0$ by requiring the background trajectory to be a geodesic of $S_{IJ}(\bm{\phi})$, and we do indeed find that this does not necessarily imply $\delta p_{\rm{nad}} = 0$.  By additionally imposing $\delta p_{\rm{nad}}=0$, we also demonstrate the possibility that both $\zeta$ and $\tilde{\zeta}$ are conserved but still with $|\zeta|\neq|\tilde{\zeta}|$.  As such, a scenario such as that depicted in Fig.\ref{example} may be possible.  In fact, in our particular model one of the fields is taken to be non-dynamical at background level, meaning that we have a situation somewhat similar to the curvaton model \cite{Moroi}\cite{Lyth+Wands}.  There is, however, a key difference.  In the curvaton model the extra degree of freedom is subdominant during inflation, meaning that it only contributes to the curvature perturbation once it comes to dominate after the end of inflation.  In our case, due to the non-minimal coupling of the extra degree of freedom, it contributes to the curvature perturbation throughout the evolution, and it is this contribution that is responsible for the non-equivalence of the curvature perturbation in the Jordan and Einstein frames.  

This paper is organised as follows. In Sec.\ref{framework} we review the general framework of cosmological perturbation theory and the geometrical approach to multi-field models of inflation. In Sec.\ref{secJF} we apply the methods outlined in Sec.\ref{framework} to the Jordan frame analysis. In Sec.\ref{efAnal} we give the analysis in the Einstein frame and compare the two different frames. In Sec.\ref{2fe} we apply our formalism to a two-field model that gives a possible example of the scenario depicted in Fig.\ref{example}. Concluding remarks and discussions are then given in Sec.\ref{conc}.          

\section{Framework and notation for background and perturbations}\label{framework}

In spite of the non-minimal coupling in \eqref{JAc}, on minimising with respect to $g^{\mu\nu}$ one is still able to recover the standard form for Einstein's equations $G_{\mu\nu} = \kappa^2T_{\mu\nu}$,\footnote{$\kappa^2 = 8\pi G = 1/M_{Pl}^2$.} but where $T_{\mu\nu}$ is now some effective energy-momentum tensor containing contributions arising from the non-minimal coupling.  As such, we can use standard perturbation methods in comparing the Jordan and Einstein frames.  In this section we clarify our notation.    

\subsection{Background equations}

At background level the universe is taken to be homogeneous and isotropic.  We further make the assumption that the spatial geometry is flat, and thus take our metric to be of the Friedmann-Lemaitre-Robertson-Walker (FLRW) form
\begin{equation}\label{backMetric}
ds^2 = -dt^2 + a(t)^2\delta_{ij}dx^idx^j\quad\quad i,j=1,~2,~3.
\end{equation}
The energy-momentum tensor is taken to be of the perfect fluid form 
\begin{equation}\label{backEnMom}
T_{\mu\nu} = pg_{\mu\nu} + (\rho + p)u_\mu u_\nu = \left(\begin{array}{cc}\rho & 0 \\0 & a^2p\,\delta_{ij}\end{array}\right),
\end{equation} 
where the energy density, momentum and scale factor, $\rho$, $p$ and $a$ respectively, are spatially independent and $u_\mu$, satisfying $u_\mu u^\mu = -1$, is the 4-velocity of the fluid.  The $00$ component and trace of Einstein's equations then give us
\begin{align}\label{backEin}
3H^2 = \kappa^2 \rho      \mbox{\,\,\,\,\,\,and\,\,\,\,\,\,}  2\dot{H} + 3H^2 = -\kappa^2 p,
\end{align}
where $H = \dot{a}/a$, and the energy-momentum constraints, $\nabla_\mu T^\mu{}_\nu=0$, additionally give us
\begin{equation}\label{backEnMom2}
\dot{\rho} + 3H(\rho + p) = 0.
\end{equation}
By working with a general form for $T_{\mu\nu}$ at this stage, all that will remain to be done when considering our specific model later on is to determine explicit expression for $\rho$ and $p$. 

\subsection{First-order perturbations}
Following the notation of \cite{KodSas}, we take our perturbed metric to be of the form
\begin{equation}\label{metDecomp}
ds^2 = -(1+2AY)dt^2 -2aBY_i dtdx^i +a^2\left[\left(1+2\mathcal{R}\right)\delta_{ij}+2H_T\frac{1}{k^2}Y_{,ij}\right]dx^idx^j,
\end{equation}
where here we only consider scalar modes.  We have already decomposed perturbations into Fourier modes with comoving wavenumber $k$ using the scalar harmonic functions $Y$ and their derivatives ($k$ labels have been suppressed), where $Y$ satisfies $(\nabla^2+k^2)Y=0$, $Y_i = -k^{-1}Y_{,i}$ and $Y_{,i} = \partial_i Y$.  The reason for not including vector and tensor modes here is that they turn out to be invariant under conformal transformations.  

Similarly, we are able to decompose the energy-momentum tensor perturbations as 
\begin{align}\nonumber\label{genFormEnMomTenPert}
\delta T_{00}& = -\rho \delta g_{00} + \delta \rho Y,\\
\delta T_{0i} = \delta T_{i0} &= p\delta g_{0i}-\delta q Y_{,i}\quad\mbox{and}\\\nonumber
\delta T_{ij} = \delta T_{ji} &= p\delta g_{ij} + a^2\left(\delta p Y \delta_{ij} + p\Pi_TY_{ij}\right),
\end{align}   
where $\Pi_T$ is the anisotropic stress perturbation and $\delta q = -(\rho + p)\delta u/k$, where $\delta u$ is the fluid velocity potential perturbation.  From Einstein's equations we then get the four explicitly gauge-invariant relations 
\bea\label{genEinEq}
&&3H\big(H\Psi-\dot{\Phi}\big)
-\frac{k^2}{a^2}\Phi
=-\frac{\kappa^2}{2}\delta_L\rho,
\nonumber\\
&&H\Psi-\dot{\Phi}
=-\frac{\kappa^2}{2}\delta_L q,
\nonumber\\
&&
H\dot{
\Psi}
+\big(2\dot{H}+3H^2\big)
\Psi
-\ddot{
\Phi}
-3H\dot{
\Phi}
=\frac{\kappa^2}{2}
\Big(\delta_L p-\frac{2}{3}p\Pi_T\Big)
\nonumber\\\mbox{and}\quad
&&
-\frac{k^2}{a^2}
\big(\Psi+\Phi\big)
=\kappa^2 
 p\Pi_T,
\eea  
where $\Psi$ and $\Phi$ are the gauge-invariant Bardeen potentials defined as
\begin{equation}\label{bardeenFields}
\Psi = A - \frac{a}{k}\left(H\sigma_g + \dot{\sigma}_g\right) \mbox{\,\,\,\,\,and\,\,\,\,\,}\Phi =  \mathcal{R} -\frac{aH}{k}\sigma_g,
\end{equation}
with $\sigma_g = a\dot{H}_T/k-B$, and $\delta_L \rho$, $\delta_L p$ and $\delta_L q$ are the gauge-invariant quantities defined as
\begin{equation}
\delta_L \rho=\delta\rho -\dot{\rho}\frac{a}{k}\sigma_g,\quad
\delta_L p=\delta p -\dot{p}\frac{a}{k}\sigma_g\quad\mbox{and}\quad
\delta_L q= \delta q+\frac{a}{k} (\rho+p)\sigma_g.
\end{equation}
From the energy-momentum constraint equations we obtain
\bea\nonumber
&&\dot{\delta_L\rho}+3H\big(\delta_L\rho+\delta_L p\big)
=\frac{k^2}{a^2}
\delta _Lq
-3(\rho+p)\dot{\Phi}\\\mbox{and}\quad
&&
\delta_L p -\frac{2}{3}p\Pi_T + \dot{\delta_L q} + 3H\delta _Lq + \Psi(\rho + p)=0.
\eea
It is worth noting here that by combining the first two expressions in \eqref{genEinEq} we obtain the Poisson equation
\bea
\frac{k^2}{a^2}
\Phi
=\frac{\kappa^2}{2}\delta\rho_m
\label{poisson},
\eea
where $\delta \rho_m:= \delta\rho-3H\delta q$.  This gives us the completely general result that $\delta\rho_m\approx 0$ on super-horizon scales ($k\ll aH$).  Once again, when considering our specific model in the following sections, all that remains is to determine explicit expressions for quantities such as $\delta\rho$, $\delta p$ and so on.     

\subsection{Curvature perturbation and its non-conservation}

As discussed in the introduction, we are interested in the gauge-invariant curvature perturbation on hypersurfaces of constant density $\zeta$ defined as 

\begin{equation}\label{curvDef}
\zeta \equiv \mathcal{R} - \frac{H}{\dot{\rho}}\delta\rho.
\end{equation} 
Taking the time-derivative of \eqref{curvDef} and making use of the energy-constraint equation to substitute for $\dot{\delta \rho}$, one finds
\begin{equation}\label{nonCons}
\dot{\zeta} = -\frac{H}{\rho + p}\delta p_{\rm{nad}} + \frac{k^2}{3}\left(\frac{\delta q}{a^2(\rho + p)}+ \sigma_g\right),
\end{equation}
so that on super-horizon scales ($k\ll aH$) this reduces to 
\begin{equation}\label{dPnadExp}
\dot{\zeta} \approx  -\frac{H}{\rho + p} \delta p_{\rm{nad}},
\end{equation}
with $\delta p _{\rm{nad}}$ as defined in \eqref{nadPdef}.  In the single-field, minimally coupled case one finds $\delta p_{\rm nad}
=-\frac{2V_{\phi}}{3H\dot{\phi}}\delta\rho_m$, thus giving $\dot{\zeta}\approx 0$ on super-horizon scales.

\subsection{A geometric approach}

As mentioned in the introduction, we can interpret the non-canonical kinetic term $G_{IJ}(\bm{\phi})$ in \eqref{JAc} as inducing a non-flat field-space with $G_{IJ}(\bm{\phi})$ as its metric.  As such, following the likes of \cite{SasakiStewart}, \cite{SasakiTanaka}, \cite{Groot}, \cite{Langlois} and \cite{Tegmark}, it is nice to use the language of manifold geometry to express equations of motion in a covariant form.  

If we consider $\phi^I(t)$ as a path on the manifold parameterised by $t$, then $\dot{\phi}^I(t)$ defines a contravariant vector at each point along the trajectory.  Associated with the field space we can define the derivative acting on a contravariant vector $X^I$ as
\begin{equation}
D X^I  = d\phi^J\nabla_J X^I = dX^I + \Gamma^I_{JK}d\phi^JX^K,
\end{equation}
where $\Gamma^I_{JK}$ and $\nabla_J$ are the connection and covariant derivative associated with the metric $G_{IJ}(\bm{\phi})$.  As an example, acting on $\dot{\phi}^I$ we have 
\begin{equation}
\frac{D \dot{\phi}^I}{dt}  = \frac{d^2\phi^I}{dt^2} + \Gamma^I_{JK}\dot{\phi}^J\dot{\phi}^K.
\end{equation}

\begin{figure}[h!]
\centering
\includegraphics[width=0.6\textwidth]{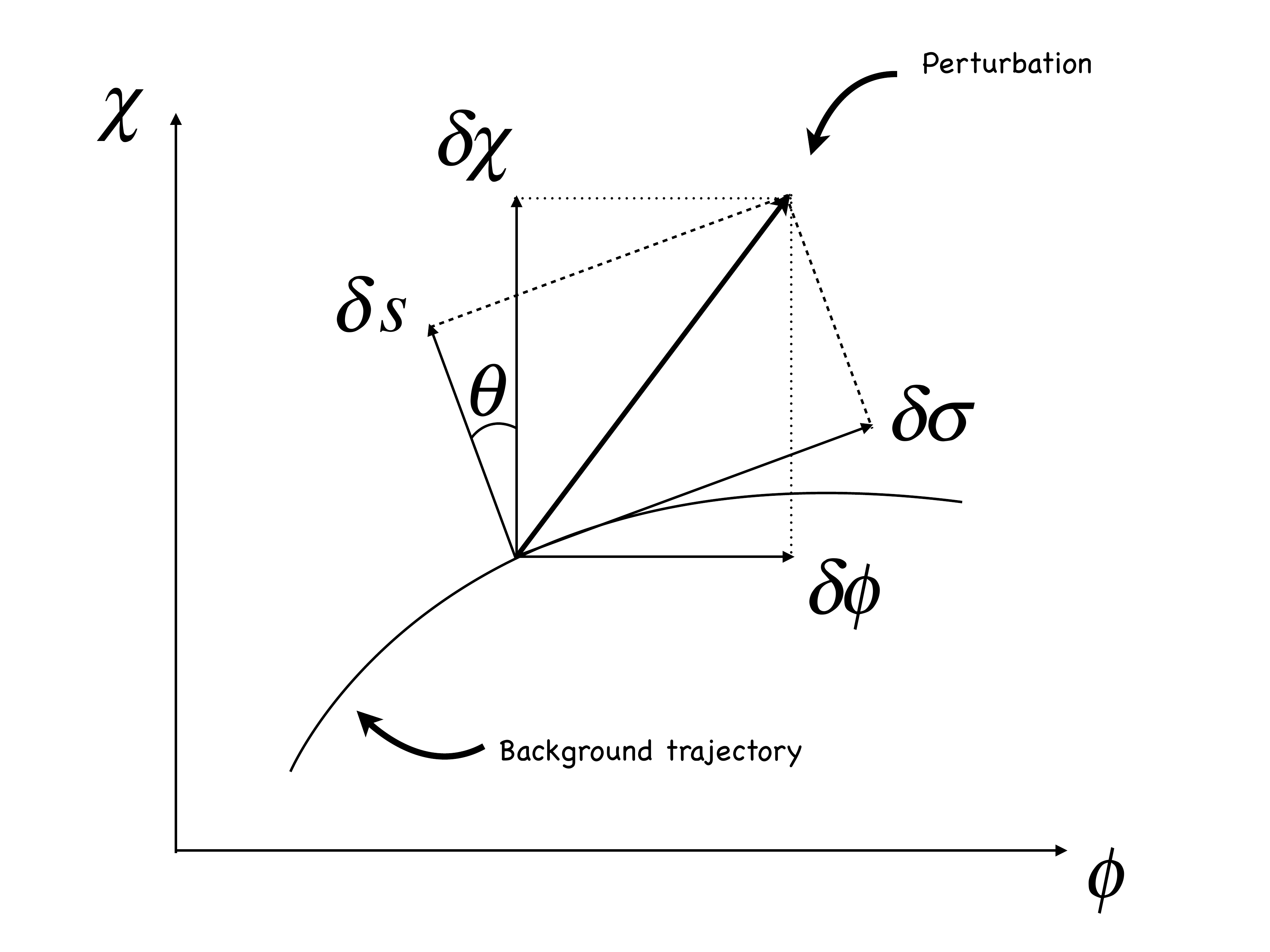}
\caption{(Taken from \cite{Gordon})  Perturbations on a background trajectory can either be decomposed in terms of the original two fields of the two-field model, $\phi$ and $\chi$, or in terms of the instantaneous adiabatic and isocurvature fields along and perpendicular to the background trajectory, $\sigma$ and $s$ respectively.  The two sets of basis vectors are related by a rotation of angle $\theta$, and in the case of a straight trajectory we have $\dot{\theta}=0$.}
\label{phasediag}
\end{figure}

As well as allowing us to write equations of motion in a compact form, this geometric interpretation also has some use when it comes to understanding adiabatic and non-adiabatic (or entropy) perturbations.  As introduced by Sasaki and Tanaka \cite{SasakiTanaka}, when considering a multi-field model of inflation with some background trajectory in field-space, one can decompose perturbations at any instant into components along and perpendicular to the background trajectory.\footnote{See \cite{Mazumdar} for an alternative decomposition recently suggested, where, in the context of the $\delta N$ formalism, perturbations are decomposed into components along the trajectory and along the hypersurface of constant e-folding number.}    These are referred to as the instantaneous adiabatic and isocurvature perturbations respectively \cite{Gordon}.  By way of example, let us recall the two-field example of \cite{Gordon}, where both fields are minimally coupled and the field-space is flat, i.e. $f(\bm{\phi})=1/2\kappa^2$ and $G_{IJ}(\bm{\phi})=\delta_{IJ}$.  Perturbations can then be decomposed into an adiabatic component along the background trajectory, $\delta \sigma$, and an isocurvature component perpendicular to the trajectory, $\delta s$, as shown in Fig.\ref{phasediag}.  On making this decomposition one finds that $\delta p_{\rm{nad}}\propto\dot{\theta}\delta s$, meaning that there is a direct correspondence between the non-adiabitic pressure perturbation and the instantaneous isocurvature field perturbation.  It is the isocurvature perturbation that sources the curvature perturbation, but only when there is a turn in the trajectory.  This concept can be extended to the case of more than two fields and a non-flat field-space, where ``perpendicular" is now defined with respect to the field-space metric.  For minimally coupled fields we find the general result that isocurvature perturbations source the curvature perturbation except when the background trajectory is a geodesic of the field-space \cite{Langlois}.  This result will be relevant when we consider the Einstein frame formulation.  Note that as the natural field-space metrics in the Jordan and Einstein frames are not the same, $G_{IJ}$ and $S_{IJ}$ respectively, the definition of isocurvature modes will also depend on the frame.  

In the context of this geometric approach, it is perhaps clear why $\zeta$ is conserved in the single-field case, as for a single-field model we cannot define a direction perpendicular to the background trajectory, and therefore have no isocurvature perturbation to act as a source.     

\section{Jordan frame analysis}\label{secJF}

In this section we apply the method of the preceding section to the action \eqref{JAc}.  The results are very similar to those given in \cite{kt}, except that here we explicitly keep the non-flat field space.  As such, the explicit expressions for quantities such as $\rho$, $p$, $\delta\rho$, $\delta p$ and $\delta q$ are given in Appendix \ref{jFeomApp}, with only the key results mentioned here.   

From \eqref{JAc} we find the effective energy-momentum tensor
to be given by
\bea\label{JFEMT}
T_{\mu\nu}
&=\frac{1}{2\kappa^2 f}
\Big[
G_{IJ}\nabla_{\mu}\phi^I
           \nabla_{\nu}\phi^J
-g_{\mu\nu}
\left(
\frac{1}{2}G_{KL}g^{\rho\sigma}
\nabla_{\rho}\phi^K
\nabla_{\sigma}\phi^L
+V
\right)+2\nabla_{\mu}\nabla_{\nu}f
-2g_{\mu\nu}\Box f
\Big],
\eea
where the last two terms are the additional contributions from the non-minimal coupling.  Note that we now drop the arguments of $f(\bm{\phi})$, $G_{IJ}(\bm{\phi})$ and $V(\bm{\phi})$.  

From the energy-momentum constraint equation $\nabla_{\nu}
T^{\nu}{}_{\mu}=0$, or by varying the action explicitly with respect to the fields $\phi^I$, we also obtain the equations of motion for the scalar fields
\bea\label{JFeom}
G_{IJ}\Box \phi^J + \Gamma_{JK|I}g^{\mu\nu}\nabla_\mu\phi^J\nabla_\nu\phi^K-V_{I}+f_I R=0,
\eea
where $R=6\big(\dot{H}+2H^2\big)$ and $X_I\equiv\frac{\partial X}{\partial \phi^I}$, except in the case of $G_{IJ}$, whose derivatives we denote as $G_{IJ,K}$,  such that 
\begin{equation}
\Gamma_{IJ|K} = \frac{1}{2}\left(G_{IK,J}+G_{JK,I}-G_{IJ,K}\right).
\end{equation}
Note that we will take our field-space to be torsion-free, i.e. $G_{IJ}=G_{JI}$.

%
At background level, the equations of motion can be written in a covariant way as 
\begin{equation}\label{jFBCovEom}
\frac{D\dot{\phi^I}}{dt} +3H\dot{\phi}^I + G^{IJ}\left(V_J - f_JR\right) = 0,
\end{equation}
where we have assumed the inverse of $G_{IJ}$, $G^{IJ}$, to exist.  
%
%
The equations of motion at first-order are given as
\begin{align}
\label{jF1CovEom}\nonumber
\frac{D^2\delta\phi^I}{dt^2}+3H\frac{D\delta\phi^I}{dt} + \frac{k^2}{a^2}\delta\phi^I& + G^{IJ}\nabla_J(\nabla_K(V-fR))\delta\phi^K-R^I{}_{JKL}\dot{\phi}^J\dot{\phi}^K\delta\phi^L
\\
&= -2G^{IJ}V_J A
+\dot{\phi}^I
\big(\dot{A}-3\dot{
\mathcal{R}}+\frac{k}{a}\sigma_g\big)
+G^{IJ}f_J
\big(
2R
A+\delta R
\big),
\end{align}
where $R^I{}_{JKL}$ is the Riemann tensor associated with the field-space and
\begin{equation}
\delta R
=6\ddot{\mathcal{R}}
-6H\big(\dot{A}
-4\dot{\mathcal{R}}\big)
-12\big(\dot{H}+2H^2\big)
 A
+\frac{2k^2}{a^2}\big(
A+2\mathcal{R}
-\frac{a}{k} (\dot{\sigma}_g+3H\sigma_g)
\big).
\end{equation}
In their current form, the equations of motion \eqref{jF1CovEom} have coupling between the scalar-field and gravitational perturbations.  However, by taking the flat gauge ($\mathcal{R} = 0$) and making use of Einstein's equations one can decouple the two, and in a covariant form the equations of motion then take the form
\begin{equation}\label{jFeomForm}
M_{2}{}^I{}_J\frac{D^2\delta\phi^J}{dt^2} + M_1{}^I{}_J\frac{D\delta\phi^J}{dt} + M_0{}^I{}_J\delta\phi^J = 0,
\end{equation}
i.e. there is mixing not only in the linear term $M_0{}^I{}_J\delta \phi^J$, but also in the derivative terms.  One can of course eliminate the mixing in the second-order derivative term by contracting with $M_2^{-1}$, but nevertheless, the additional mixing makes solving the equations of motion much more involved than in the standard minimally coupled case.  

Explicit expressions for $M_2$, $M_1$ and $M_0$ are given in Appendix \ref{jFeomApp}, and as a check one can demonstrate that the known result is recovered for the minimally coupled case, namely (see e.g. \cite{Tegmark})
\begin{align}\label{covDecEomMinCoup}
&\frac{D^2\bm{\delta\phi}}{dt^2}+3H\frac{D\bm{\delta\phi}}{dt} + \frac{k^2}{a^2}\bm{\delta\phi}\\\nonumber
&\quad\quad=\left[-\bm{\nabla^\dagger\nabla}V+(3-\frac{\dot{H}}{H^2})\dot{\bm{\phi}}\dot{\bm{\phi}}^\dagger + \frac{1}{H}\dot{\bm{\phi}}\frac{D\dot{\bm{\phi}}^\dagger}{dt}+\frac{1}{H}\frac{D\dot{\bm{\phi}}}{dt}\dot{\bm{\phi}}^\dagger+\bm{R}(\dot{\bm{\phi}},\dot{\bm{\phi}})\right]\bm{\delta\phi},
\end{align}
where an index-free notation has now been adopted (see Appendix \ref{jFeomApp} for details).

\subsection{Non-conservation of the curvature perturbation}

We first introduce the two gauge invariant variables
\bea\label{JHdefs}
{\cal K}^{JK}
:=\delta\phi^J \dot{\phi}^K-\delta\phi^K \dot{\phi}^J
\quad\mbox{and}\quad
{\cal H}^{IJ}
:= \ddot{\phi}^J\delta\phi^I
-\dot{\phi}^I\dot{\delta\phi}^J
+A\dot{\phi}^I\dot{\phi}^J,
\eea
where the former corresponds to the entropy perturbation between the $I$'th and $J$'th fields.  Note that when we say an effectively single-field adiabatic limit is reached, this implies that $\delta\phi^I\propto\dot{\phi}^I$, meaning that $\mathcal{K}^{IJ}=0$ for all $I$, $J$.  Also note that we are no longer taking the flat gauge $\mathcal{R} = 0$.  Using these two new variables we find that we are able to write the non-adiabatic pressure \eqref{nadPdef} in the form 
%
\begin{equation}\label{dPnadFormJF}
\delta p_{\rm{nad}} = \mathcal{N}_{IJ}\mathcal{K}^{IJ} + \mathcal{P}_{IJ}\dot{\mathcal{K}}^{IJ} + \mathcal{Q}_{IJ}\mathcal{H}^{IJ} + \mathcal{T}_{IJ}\dot{\mathcal{H}}^{IJ}
\end{equation}
(see Appendix \ref{jFeomApp} for explicit expressions).  The full derivation of this new form is too lengthy to be included here, but we do note that the relation $\delta\rho_m \approx 0$ played a key role.  In terms of $\mathcal{K}^{IJ}$ and $\mathcal{H}^{IJ}$, $\delta\rho_m \approx 0$ can be expressed as
\bea
\Big(G_{IJ}+\frac{3f_If_J}{f}\Big)
{\cal H}^{IJ}
+G_{IJ,K}\dot{\phi}^I\mathcal{K}^{JK}
\approx
\frac{3f_K}{2f}
\big(G_{IJ}+2f_{IJ}\big)
\dot{\phi}^I
{\cal K}^{JK},
\label{cons}
\eea
which we will use again below.  

In general it is rather difficult to interpret \eqref{dPnadFormJF}.  In particular, there seems to be no geometrical interpretation as to when $\delta p_{\rm{nad}} = 0$, i.e. when the curvature perturbation is conserved.  We can, however, relatively easily recover the known result that the curvature perturbation is conserved in the single-field case, despite the non-minimal coupling \cite{Naruko}.  In the single-field case we know that $\mathcal{K}^{\phi\phi}=0$, which means that the only remaining contributions to $\delta p_{nad}$ are proportional to $\mathcal{H}^{\phi\phi}$ and its derivative.  However, in the single-field case and on super-horizon scales, \eqref{cons} reduces to 
\begin{equation}\label{singFieldDeltaRhom}
\left(G_{\phi\phi}+\frac{3f_\phi^2}{f}\right)\mathcal{H}^{\phi\phi} \approx 0,
\end{equation}
which tells us that unless $G_{\phi\phi}+3f^2_\phi/f = 0$,\footnote{We shall see that the case $G_{\phi\phi}+3f^2_\phi/f = 0$ corresponds to the field having vanishing kinetic term in the Einstein frame.} then we must have $\mathcal{H}^{\phi\phi} = 0$ and thus the curvature perturbation is conserved even in the non-minimally coupled case.

We note that in the minimally coupled case, where $f={\rm const}$, \eqref{dPnadFormJF} gives
\bea
\delta p_{\rm nad}
&\approx&
\frac{2V_K}
  {G_{LM}\dot{\phi}^L\dot{\phi}^M}
G_{IJ}\dot{\phi}^I {\cal K}^{JK},
\eea  
which, on using the background equations of motion \eqref{jFBCovEom}, can be re-expressed as 
\begin{equation}\label{comFormdP}
\delta p_{\rm{nad}} \approx  -\frac{2G_{IJ}\dot{\phi}^I}
  {G_{LM}\dot{\phi}^L\dot{\phi}^M}
G_{KN} \frac{D\dot{\phi^N}}{dt}{\cal K}^{JK},
\end{equation}
meaning that the curvature perturbation will be conserved on super-horizon scales if the background trajectory follows a geodesic of the field-space (see e.g. \cite{Langlois}).

\section{Einstein frame analysis and comparing the frames}\label{efAnal}

In this section we make a transformation into the Einstein frame, again apply the methods of Sec.\ref{framework} and compare the results with the Jordan frame analysis of the preceding section.  Consider our original metric to be given in terms of some new metric $\tilde{g}_{\mu\nu}$ as 
\begin{equation} 
g_{\mu\nu} = \Omega\tilde{g}_{\mu\nu}.
\end{equation}
Under this conformal transformation, 
taking $\Omega = \frac{1}{2\kappa^2f}$ we can obtain an action whose gravitational part is of the canonical Einstein-Hilbert form, namely
\begin{equation}\label{eFAction}
S = \int d^4x\sqrt{-\tilde{g}}\left\{\frac{\tilde{R}}{2\kappa^2} - \frac{1}{2}S_{IJ} \tilde{g}^{\mu\nu}\tilde{\nabla}_\mu \phi^I \tilde{\nabla}_\nu \phi^J - \frac{1}{(2\kappa^2f)^2}V\right\},
\end{equation}
where, comparing with \eqref{EAc}, we have $\tilde{V} = V/(2\kappa^2f)^2$ and we have defined the new quantity\footnote{Note that we require the eigenvalues of the matrix $S_{IJ}$
to be positive in order to avoid the appearance of ghosts in our model.}
\begin{equation} 
S_{IJ} = \frac{1}{2\kappa^2 f}\left[G_{IJ} + 3\frac{f_If_J}{f}\right].
\end{equation}

Ideally we would like to be able to bring the action (\ref{eFAction}) into a form where the kinetic term is diagonal in the field space.  However, in general this will not be possible \cite{kaiser2}, so we concede to proceeding with \eqref{eFAction} as it is.  In the geometric interpretation, we see that in the Einstein frame we have a new induced field-space with metric $S_{IJ}$.  We are thus able to define a connection and covariant derivative associated with this metric in the exact same way as we did for $G_{IJ}$.  As an aside, note that whilst the field-space metrics in the Jordan and Einstein frames are different, the general form of the non-canonical kinetic term is the same in both frames, i.e. both can be interpreted as inducing a non-flat field-space.  However, if we had taken a more complex form of kinetic term in the Jordan frame, such as that of Dirac-Born-Infeld inflation models, then we would find that this form is not preserved under the conformal transformation to the Einstein frame \cite{Easson}.   

The form of action \eqref{eFAction} is something much more familiar to us (see e.g. \cite{Tegmark}).  As such, we once again defer details of the analysis to Appendix \ref{EFApp}, noting only the key results below.    

On varying the matter action with respect to $\tilde{g}^{\mu\nu}$ we find
\begin{align}\label{fullEMTen}
\tilde{T}_{\mu\nu} = \tilde{g}_{\mu\nu}\left\{ - \frac{1}{2}S_{IJ}\tilde{g}^{\alpha\beta}\tilde{\nabla}_\alpha \phi^I \tilde{\nabla}_\beta \phi^J - \frac{V}{(2\kappa^2f)^2}\right\} +S_{IJ}\tilde{\nabla}_\mu \phi^I \tilde{\nabla}_\nu \phi^J, 
\end{align}
and the equations of motion for the $N$ fields are found to be
\begin{align}\label{fullEOM}
S_{IJ}\tilde{\Box}\phi^J + \Gamma^{{}^{(S)}}_{JK|I}\tilde{g}^{\mu\nu}\tilde{\nabla}_\mu\phi^J\tilde{\nabla}_\nu\phi^K + \frac{2V}{(2\kappa^2)^2f^3}f_I - \frac{1}{(2\kappa^2f)^2}V_I = 0.
\end{align}
%
%
At background level these reduce to
\begin{equation}\label{efBackEom}
\frac{D^{{}^{(S)}}\phi^{\prime I}}{d\tilde{t}}+3\tilde{H}\phi^{\prime I} +\frac{1}{(2\kappa^2f)^2}S^{IJ}\left(V_J-\frac{2V}{f}f_J\right)=0,
\end{equation} 
where we assume $S_{IJ}$ to have an inverse, $S^{IJ}$, a prime denotes derivatives with respect to the time $\tilde{t}$, which we will define shortly, and a superscript ``${}^{(S)}$" is used to distinguish field-space objects associated with the metric $S_{IJ}$ as opposed to $G_{IJ}$.

In the usual fashion, the perturbed equations of motion for the scalar fields can be decoupled from gravitational perturbations by taking the flat gauge ($\tilde{\mathcal{R}} = 0$).  They take the same form as in \eqref{covDecEomMinCoup}, but with all covariant derivatives and other field-space quantities now understood to be those associated with the metric $S_{IJ}$ and field perturbations now being those in the flat-gauge as defined in the Einstein frame as opposed to the Jordan frame, i.e. $\tilde{\mathcal{R}} = 0$ not $\mathcal{R} = 0$.   
%
%

\subsection{Curvature perturbation and its non-conservation}

On inserting the appropriate results for $\delta\tilde{\rho}$, $\delta \tilde{p}$, $\tilde{\rho}^\prime$ and $\tilde{p}^\prime$ (as given in Appendix \ref{EFApp}) into \eqref{nadPdef} one finds 
\begin{equation}\label{pNad}
\delta \tilde{p}_{nad} = -\frac{1}{(2\kappa^2 f)^2}\left\{\frac{2V_I\phi^{\prime I}}{3\tilde{H}(\tilde{\rho} + \tilde{p})}\delta\tilde{\rho}_m + 2V_I\tilde{\Delta}^I - \frac{4Vf^\prime\delta\tilde{\rho}}{3\tilde{H}f(\tilde{\rho} + \tilde{p})} - \frac{4V\delta f}{f}\right\},
\end{equation}
where $\delta \tilde{\rho}_m := \delta\tilde{\rho} - 3\tilde{H}\delta\tilde{q}$ and $\tilde{\Delta}^I$ is as defined for the Jordan frame in \eqref{sFpCg}, but with Jordan frame quantities replaced with the equivalent Einstein frame ones.

In the super-horizon limit, i.e. $k \ll \tilde{a}\tilde{H}$, \eqref{pNad} can be shown to reduce to
\begin{align}\label{lSPnad}
\delta \tilde{p}_{nad}  = \frac{2S_{IJ}\phi^{\prime I}}{(2\kappa^2f)^2(\tilde{\rho} + \tilde{p})}(V_K - \frac{2V}{f}f_K)\tilde{\mathcal{K}}^{JK},
\end{align}
and using the background equations of motion this can then be re-expressed as 
\begin{align}\label{eFdPnad}
\delta \tilde{p}_{nad}  = -\frac{2S_{IJ}\phi^{\prime I}}{(2\kappa^2f)^2(\tilde{\rho} + \tilde{p})}S_{KL}\frac{D^{(S)}\phi^{\prime L}}{d\tilde{t}}\tilde{\mathcal{K}}^{JK}.
\end{align}
This is, of course, of the same form as \eqref{comFormdP} with the substitution $G_{IJ} \rightarrow S_{IJ}$, and so we see that in the case that the trajectory follows a geodesic in the field space, i.e. $\frac{D^{(S)}\phi^{\prime L}}{d\tilde{t}}=0$, the curvature perturbation will be conserved.  This is the natural extension of the well-known result of Gordon {\it et al} discussed in Sec.\ref{framework}, where in a flat field-space a straight trajectory led to conservation of the curvature perturbation.  

\subsection{Comparing the frames}

In both the Jordan and Einstein frames we are able to make the metric decomposition (\ref{metDecomp}), but since the two metrics are related as $g_{\mu\nu} = \Omega \tilde{g}_{\mu\nu}$, these two decompositions are not independent.  Starting with the background metric, we have 
\begin{equation}\label{compBack}
ds^2 = -dt^2 +a^2(t)\delta _{ij}dx^idx^j = \Omega_0 d\tilde{s}^2 = \Omega_0\left(-d\tilde{t}^2 + \tilde{a}^2(\tilde{t})\delta_{ij}d\tilde{x}^id\tilde{x}^j\right),
\end{equation}
where $\Omega_0$ indicates the background value of the conformal factor.  This gives us
\bea\label{backRel}
\tilde a=\frac{a}{\sqrt{\Omega_0}},\quad
d{\tilde t}=\frac{dt}{\sqrt{\Omega_0}},\quad
d\tilde{x}^i = dx^i \quad \mbox{and}\quad{\tilde H}
=\sqrt{\Omega_0}
 \Big(
 H-\frac{\dot{\Omega}_0}{2\Omega_0}
 \Big),
\eea
so we can see that even at background level there are many apparent non-equivalences between the two frames.  For example, the notion of an accelerating expansion is different in the two frames \cite{Piao}\cite{Qiu}, $\dot{H}/H^2\neq\tilde{H}^\prime/\tilde{H}^2$, so that the notion of slow-roll is not equivalent and $aH\neq\tilde{a}\tilde{H}$, meaning that the idea of super-horizon scales is also not equivalent.  

Using the background relations \eqref{backRel} we can then establish that for the perturbations we have
\bea\label{sJFEFrel}
\tilde A=A-\frac{\delta\Omega}{2\Omega_0},\quad
\tilde {\mathcal{R}}=\mathcal{R}-\frac{\delta\Omega}{2\Omega_0},\quad
\tilde B=B\quad\mbox{and}\quad
\tilde H_T=H_T.
\eea 
As such, we can see that if we are able to take a gauge such that $\delta\Omega=0$ then the perturbations are equivalent, despite the apparent non-equivalence at background level.  

In the single-field case $\delta\Omega \propto \delta\phi$, so that $\delta\Omega = 0$ corresponds to the constant-field and comoving ($\delta T^0{}_{i}=0$) gauges.  As such, the comoving curvature perturbation is equivalent in the two frames \cite{Gong}\cite{Kubota}.  Using the fact that on super-horizon scales the comoving and constant-density curvature perturbations coincide, we can therefore conclude that the latter is also equivalent in the two frames.  

In the more general case, however, we may not be able to choose a gauge in which $\delta\Omega =0$, and even when this is possible there is no guarantee that this choice of gauge will coincide with the comoving or constant energy-density gauges.  In comparing $\zeta$ and $\tilde{\zeta}$ in the more general case, we in fact choose to work with the curvatures on comoving hypersurfaces, $\mathcal{R}_c$ and $\tilde{\mathcal{R}}_c$, defined as
\bea
\tilde{\cal R}_c=
\tilde{\mathcal{R}}
+\frac{{\tilde H} \delta \tilde{q}}
       {{\tilde \rho}+{\tilde p}}\quad\mbox{and}
\quad
{\cal R}_c=
\mathcal{R}
+\frac{{H}{\delta q}}
       {{\rho}+{ p}},
\eea
where $\delta q$ and $\delta \tilde{q}$ are as defined in \eqref{genFormEnMomTenPert} for the Jordan and Einstein frames respectively (see \eqref{expPertQuantJF} and \eqref{EEdE} for explicit expressions).  This is because on super horizon scales we have $\mathcal{R}_c \approx \zeta$ and $\tilde{\mathcal{R}}_c\approx\tilde{\zeta}$, and the expressions for $\mathcal{R}_c$ and $\tilde{\mathcal{R}}_c$ turn out to be simpler than those for $\zeta$ and $\tilde{\zeta}$.
Taking their difference we have 
\bea
\zeta - \tilde{\zeta} \approx {\cal R}_c-\tilde {\cal R}_c
&=&
-\frac{\delta f}{2f}
+\frac{{H}{\delta q}}
       {{\rho}+{ p}}
-\frac{{\tilde H}{\tilde \delta q}}
       {{\tilde \rho}+{\tilde p}}.
\eea
Thanks to \eqref{sJFEFrel}, taking the longitudinal gauge in one frame is equivalent to taking it in the other.\footnote{See \cite{Brown} for a discussion on the relation between gauge choices made in the Jordan and Einstein frames.}  Thus, on taking the longitudinal gauge, and after some manipulation, we obtain
\bea\nonumber
\zeta - \tilde{\zeta}
&\approx&
\frac{f_KS_{IJ}\dot{\phi}^I{\cal K}^{JK}}
{2f(S_{MN}\dot{\phi}^M\dot{\phi}^N)}
+\frac{H S_{IJ}\dot{\phi}^I\Big\{
{\cal K}^{JK}
\Big(G_{LK}\dot{\phi}^L+2f_{KL}\dot{\phi}^L
-2Hf_K\Big)
+2f_{K}{\cal H}^{JK}
\Big\}}{(S_{MN}\dot{\phi}^M\dot{\phi}^N)
(G_{PQ}\dot{\phi}^P\dot{\phi}^Q+2(\ddot{f}-H\dot{f}))}.
\eea
With the help of \eqref{cons}, one can then re-express the term of the form $2HS_{IJ}\dot{\phi}^If_K\mathcal{H}^{JK}$ in terms of $\mathcal{K}^{IJ}$, leading finally to  
\begin{equation}\label{genDiff}
\zeta - \tilde{\zeta}
\approx \mathcal{A}_{JK}\mathcal{K}^{JK} + \mathcal{B}_{JK}\dot{\mathcal{K}}^{JK},
\end{equation}
where
\begin{align}\nonumber
\mathcal{A}_{JK} &= \frac{1}{\mathcal{C}}\Bigg\{\Bigg[\Bigg(\frac{G_{PQ}\dot{\phi}^P\dot{\phi}^Q+2\left(\ddot{f}-H\dot{f}\right)}{2f}-2H^2\Bigg)f_KG_{IJ} \\\label{genDiffC1} &\quad\quad\quad\quad+ 2H\dot{\phi}^Lf_{KL}G_{IJ}-2H\dot{f}G_{IJ,K}\Bigg]\dot{\phi}^I-2Hf_KG_{IJ}\ddot{\phi}^I\Bigg\},\\\label{genGiffC2}
\mathcal{B}_{JK} &= \frac{2Hf_KG_{IJ}\dot{\phi}^I}{\mathcal{C}}\quad\quad\mbox{and}\\\label{genDiffC3}
\mathcal{C} &= 2\kappa^2fS_{MN}\dot{\phi}^M\dot{\phi}^N\left(G_{PQ}\dot{\phi}^P\dot{\phi}^Q+2\left(\ddot{f}-H\dot{f}\right)\right).
\end{align}
With this expression for the difference written wholly in terms of $\mathcal{K}^{IJ}$ and its derivatives, it is explicitly clear that it is the isocurvature modes that are responsible for any discrepancy between the two frames.  In particular, we see that the difference vanishes in the single-field case and in any scenario where an effectively single-field adiabatic limit is reached.  

Note that as \eqref{genDiff} and \eqref{eFdPnad} are given wholly in terms of $\mathcal{K}^{IJ}$ and its derivatives, it should also be possible to re-write \eqref{dPnadFormJF} in a similar form.  It is therefore clear that even in the Jordan frame the non-conservation of the curvature perturbation is purely a consequence of the isocurvature perturbations, and in any effectively single-field adiabatic limit $\dot{\zeta}=0$ is recovered.  Relating to this point, we also note that as a consequence of \eqref{backRel} we have the relation $\tilde{\mathcal{K}}^{IJ} = \mathcal{K}^{IJ}/\sqrt{2\kappa^2f}$.  This means that the vanishing of the isocurvature perturbations in an effectively single-field adiabatic limit, where $\delta\phi^I \propto \dot{\phi}^I$, is independent of the frame, i.e. $\tilde{\mathcal{K}}^{IJ} = \mathcal{K}^{IJ} = 0$, in turn giving us $\dot{\tilde{\zeta}} = \dot{\zeta} = 0$ and $\tilde{\zeta} = \zeta$.  The equivalence of the curvature perturbations and their statistical properties in the absence of isocurvature modes and in the slow roll limit is discussed in \cite{Chiba2}.  

More generally, the fact that $\zeta\neq\tilde{\zeta}$ suggests that the evolution of the two curvature perturbations will also be different, which in turn means that the idea of non-adiabaticity may be frame dependent, i.e. $\delta \tilde{p}_{\rm{nad}} =0 \nLeftrightarrow \delta p_{\rm{nad}} = 0$.  It is important to stress, however, that this result simply highlights the fact that $\zeta$ itself is not an observable quantity.  Any observable predictions made should remain independent of the frame, irrespective of whether the adiabatic limit is reached,
so long as we are careful to keep track of how the non-minimal coupling effects matter, rulers and clocks in the two frames.

As a final comment, we note that until now we have been defining quantities in terms of effective fluid quantities such as $\delta\rho$ and $\delta q$.  In particular, the comoving curvature perturbation was defined as the curvature perturbation on hypersurfaces comoving with the effective fluid, i.e. $\delta u = 0 \Rightarrow \delta q = 0 = \delta T^0{}_i$.  In terms of fields, however, the natural definition of comoving is that the adiabatic component of the perturbation in field-space be zero, i.e. $G_{IJ}\dot{\phi}^I\delta\phi^J = 0$ in the Jordan frame or $S_{IJ}\phi^{\prime I}\delta\phi^J = 0$ in the Einstein frame.  In the Einstein frame, it is clear from the expression for $\delta\tilde{q}$ in \eqref{EEdE} that the two definitions are equivalent.  In the Jordan frame, however, with the expression for $\delta q$ as given in \eqref{expPertQuantJF}, the equivalence is by no means obvious.  Only in the single-field case, using \eqref{singFieldDeltaRhom}, is it easy to show that $\delta q \propto \delta \phi$, thus recovering the equivalence of the two definitions.
  
\section{A two-field curvaton-like example}\label{2fe}

In this section we attempt to find an interesting example along the lines of that mentioned in the introduction and depicted in Fig.\ref{example}.  In order to do so, we simplify to a two-field model with action
\begin{equation}
S=\int{d^4x\sqrt{-g}}\left\{f(\phi, \chi)R-\frac{1}{2}G_{\phi\phi}(\partial\phi)^2-\frac{1}{2}G_{\chi\chi}(\partial\chi)^2 -G_{\phi\chi}g^{\mu\nu}\partial_\mu\phi\partial_\nu\chi- V(\phi, \chi)\right\}.
\end{equation}
To simplify things further, we take a straight background trajectory defined by $\dot{\chi}=0$.  Applying this constraint to the equations of motion one finds that \eqref{eFdPnad} simplifies to
\begin{align}\nonumber\label{EF}
\delta \tilde{p}_{nad} =\frac{2\tilde{\mathcal{K}}^{\phi\chi}\phi^{\prime 3}}{2f(2f)^2(\tilde{\rho}+\tilde{p})}\big\{&\left(fG_{\chi\phi}+3f_\phi f_\chi\right)G_{\phi\phi,\phi} - \left(fG_{\phi\phi}+3f_\phi^2\right)\left(2G_{\chi\phi,\phi}-G_{\phi\phi,\chi}\right)\\
 &+ \left(G_{\phi\phi}+6f_{\phi\phi}\right)\left(f_\phi G_{\phi\chi}-f_\chi G_{\phi\phi}\right)\big\},
\end{align} 
and turning to the difference between the two frames, the general expression \eqref{genDiff} reduces to
\begin{equation}
\zeta - \tilde{\zeta}
= (\mathcal{A}_{\phi\chi}-\mathcal{A}_{\chi\phi})\mathcal{K}^{\phi\chi} + (\mathcal{B}_{\phi\chi}-\mathcal{B}_{\chi\phi})\dot{\mathcal{K}}^{\phi\chi}.
\end{equation}
%
%
%
Note that we will take $\kappa^2 = 1$ in this section.  With $\mathcal{K}^{\phi\chi}=-\dot{\phi}\delta\chi$, we see that in order to track the difference between the two frames we are going to need to solve the equation of motion for $\delta\chi$.  However, in general the two perturbations $\delta\chi$ and $\delta\phi$ will be coupled, so in order to simplify the situation as much as possible, we would like to try and decouple $\delta\chi$ from $\delta\phi$.  

To help us do this we consider things in the Einstein frame.  Here, we know that on super-horizon scales the isocurvature field perturbation, i.e. the part perpendicular to the background field trajectory, is in general not sourced by the adiabatic component along the field direction (see e.g. \cite{Tegmark}).  Thus, if we make the perturbation $\delta\chi$ coincide with the isocurvature perturbation then we should find that its equation of motion is decoupled from $\delta\phi$.  In the case of a flat field-space we know that $\delta\chi$ does already correspond to the isocurvature perturbation for the background trajectory $\dot{\chi}=0$, but now we must require that the isocurvature mode is perpendicular to the background trajectory with respect to the effective field-space $S_{IJ}$.  As such, requiring $\delta\chi$ to be perpendicular to the background trajectory $\dot{\chi}=0$ in fact requires that $S_{\phi\chi}=S_{\chi\phi}=0$. 
The simplest way to realise this is for us to assume the field-space metric in the Jordan frame to be diagonal, i.e. $G_{\phi\chi}=G_{\chi\phi}=0$, and either $f_\phi=0$ or $f_\chi=0$.  Let us choose the case where $f_\phi=0$,\footnote{Choosing $f_\chi = 0$ turns out not to give us the example we are after.  Also note that in the most general case we only require that $f_\phi |_{\chi = const} =0$.  However, for simplicity we make the assumption that $f = f(\chi)$.}  which is a rather interesting case, as combined with the background trajectory property $\dot{\chi}=0$ it leads to $\dot{f}=0$.  These two assumptions therefore greatly simplify many of the general expressions obtained previously, as well as allowing us to remove the sourcing of $\delta\chi$ by $\delta\phi$.  Also note that as $f$ is a constant, the Einstein and Jordan frames become equivalent at background level.  With inflation being driven by the single scalar field $\phi$, and $\chi$ being non-dynamical, our model is somewhat similar to a curvaton-like model.  However, the non-minimal coupling of the additional degree of freedom $\chi$ means that, unlike in the case of the curvaton, it does make a contribution to the curvature perturbation throughout the evolution.         

Recall that we are now looking for an example where $\delta p_{\rm{nad}}$ vanishes in one frame but not in the other.  Given that we have a better understanding of the conditions under which the curvature perturbation is conserved in the Einstein frame, we choose to require $\delta\tilde{p}_{\rm{nad}} = 0$.  In combination with the previous constraints, this gives us the condition $fG_{\phi\phi,\chi}=G_{\phi\phi}f_\chi$.\footnote{Note that due to a cancellation, in requiring $\delta\tilde{p}_{\rm{nad}}=0$ we in fact only require $D^{(S)}\chi^\prime/d\tilde{t} = 0$, with no such restriction on $D^{(S)}\phi^\prime/d\tilde{t}$.}  After satisfying all these conditions, the background equations of motion and Einstein equations in the Jordan frame reduce to 
\begin{align}
f_\chi R = V_\chi& & &\frac{D\dot{\phi}}{dt} +3H\dot{\phi} + \frac{V_\phi}{G_{\phi\phi}}=0\\
3H^2=\frac{G_{\phi\phi}\dot{\phi}^2}{4f}+\frac{V}{2f}\label{Hdot}& &&2\dot{H} =-\frac{1}{2f}G_{\phi\phi}\dot{\phi}^2
\end{align}
and the equation of motion for $\delta\phi$ and $\delta\chi$ become
\begin{align}\label{reddChiEom}
&\ddot{\delta\chi}+(3H+c_\chi)\dot{\delta\chi}+\left(\frac{k^2}{a^2}-m_\chi^2\right)\delta\chi = 0\\\label{reddPhiEom}
&\ddot{\delta\phi}+(3H+c_\phi)\dot{\delta\phi} +\left(\frac{k^2}{a^2}-m_\phi^2\right)\delta\phi = \gamma\dot{\delta\chi} + \lambda\delta\chi
\end{align}
with explicit expressions for the coefficients given in Appendix \ref{jFeomApp}.  Interestingly, we find that the condition for conservation of the curvature perturbation in the Einstein frame, $fG_{\phi\phi,\chi}=G_{\phi\phi}f_\chi$, results in the vanishing of $\delta\phi$ source terms in the $\delta\chi$ equation of motion on {\it all} scales, as opposed to the expected vanishing discussed in the previous paragraph, which is only valid on super-horizon scales.   

The expression for $\zeta - \tilde{\zeta}$ simplifies to 
\begin{align}
\zeta - \tilde{\zeta}
=\frac{2f_\chi}{G_{\phi\phi}\dot{\phi}^2}\left\{(H^2+\dot{H})\delta\chi-H\dot{\delta\chi}\right\},
\end{align}
and taking its derivative we find 
\begin{align}
\frac{\rm{d}}{\rm{dt}}(\zeta - \tilde{\zeta})& =\frac{\rm{d}}{\rm{dt}}\zeta
=-\frac{2Hf_\chi}{G_{\phi\phi}\dot{\phi}^2}\left\{\ddot{\delta\chi}-H\left(1+2\eta\right)\dot{\delta\chi} + H^2\left(2\eta\left(1-\epsilon\right)+\epsilon(2+\xi)\right)\delta\chi\right\},
\end{align}
where we have defined the slow-roll parameters
\begin{equation}\label{slowRP}
\eta = \frac{1}{H\dot{\phi}}\frac{D\dot{\phi}}{dt}\quad\quad\epsilon = -\frac{\dot{H}}{H^2}\quad\quad\mbox{and}\quad\quad\xi=\frac{\ddot{H}}{H\dot{H}}.
\end{equation}
From this result we see that despite $\tilde{\zeta}$ being conserved, $\zeta$ in general is not, meaning that adiabatic evolution in one frame does not correspond to adiabatic evolution in the other.  Let us now go a step further and consider whether we could have conservation of both curvature perturbations but still maintain a non-zero difference between them.  In this case, requiring $\frac{\rm{d}}{\rm{dt}}(\zeta - \tilde{\zeta}) = 0$ gives us a second differential equation for $\delta{\chi}$, and we need to check that this is compatible with \eqref{reddChiEom}.  We would also like to check whether or not this condition can be satisfied by an inflationary solution.  Let us simplify by taking the slow roll approximation.  Under this approximation we would like to take our background equation of motion and Friedmann equation as
\begin{align}\label{slowPhiEom}
&3HG_{\phi\phi}\dot{\phi} + V_\phi = 0\\\label{Hsquared}
&3H^2=\frac{V}{2f},
\end{align}
which amounts to $\epsilon,~\eta,~\xi \ll1$.  By taking the time-derivative of \eqref{Hsquared}, or directly from \eqref{Hdot}, we find 
\begin{equation}\label{sr1}
\epsilon \ll 1 \quad\Rightarrow\quad \frac{f}{G_{\phi\phi}}\left(\frac{V_\phi}{V}\right)^2\ll 1
\end{equation}
and also by differentiating \eqref{slowPhiEom} and $\epsilon$ we find that 
\begin{equation}\label{sr2}
\eta,~\xi \ll 1\quad \Rightarrow\quad \frac{f}{G_{\phi\phi}}\frac{V_{\phi\phi}}{V}\ll 1. 
\end{equation}
If these conditions are satisfied, then to first order in the slow-roll approximation the fractional change of $\zeta - \tilde{\zeta}$ over a Hubble time is given as 
\begin{equation}\label{fracDiff}
\frac{1}{H}\frac{\rm{d}}{\rm{dt}}\ln(\zeta - \tilde{\zeta}) =-\frac{\ddot{\delta\chi}-H\left(1+2\eta\right)\dot{\delta\chi} + 2H^2\left(\eta+\epsilon\right)\delta\chi}{(H^2+\dot{H})\delta\chi-H\dot{\delta\chi}}.
\end{equation}

Let us now try so solve \eqref{reddChiEom} for $\delta\chi$ and determine whether or not this can be compatible with \eqref{fracDiff}$=0$.  In order to do so we first introduce the variables
\begin{equation}
u_\chi = a\sqrt{G_{\chi\chi}+\frac{3f_\chi^2}{f}}\delta\chi\quad\quad\mbox{and}\quad\quad u_\phi = a\sqrt{G_{\phi\phi}}\delta\phi
\end{equation} 
and use conformal time, $dt = ad\tau$, to bring \eqref{reddChiEom} and \eqref{reddPhiEom} into the form
\begin{align}\label{quantchi}
&u_\chi^{\prime\prime}+\left(k^2-\frac{a^{\prime\prime}}{a}-a^2m_{\chi \rm{tot}}^2\right)u_\chi = 0\\\label{quantphi}
&u_\phi^{\prime\prime}+\left(k^2-\frac{a^{\prime\prime}}{a}-a^2m_{\phi\rm{tot}}^2\right)u_\phi = \alpha u_\chi^\prime + \beta u_\chi
\end{align}  
where here a prime denotes taking the derivative with respect to the conformal time and the coefficients are given in Appendix \ref{jFeomApp}.  Making the simplifying assumption $G_{\chi\chi} = 1$, such that $m_{\chi tot}^2 = m_\chi^2$, if we suppose that $m_\chi^2/H^2 \sim O(\epsilon)$ then we can solve \eqref{quantchi} explicitly, with the well-known result
\begin{equation}
u_\chi = \frac{1}{\sqrt{2k}}e^{-ik\tau}\left(1-\frac{i}{k\tau}\right),
\end{equation}
so that on super-horizon scales we have $\delta\chi \simeq const = \frac{H}{\sqrt{2k^3}}(1+3f_\chi^2/f)^{-1/2}$.\footnote{Note that from here on we neglect the phase factor $e^{-i\pi/2}$.}  Using this result we find 
\begin{equation}\label{resDiff}
\frac{1}{H}\frac{\rm{d}}{\rm{dt}}\ln(\zeta - \tilde{\zeta}) \sim 
O(\epsilon)
\quad\quad\mbox{and}\quad\quad
\zeta - \tilde{\zeta} = \frac{f_\chi}{2f\epsilon}\frac{H}{\sqrt{2k^3}}(1+3f_\chi^2/f)^{-1/2},
\end{equation}
so that to zeroth order in slow roll we also have conservation of $\zeta$ whilst maintaining a non-zero $\zeta - \tilde{\zeta}$.  Comparing \eqref{resDiff} with standard slow-roll expressions, we see that in order for $\zeta - \tilde{\zeta}$ to be of the correct order of magnitude we require $f_\chi/\sqrt{f}\sim O(\epsilon^{1/2})$.

We would now like to establish whether or not we can achieve the condition $m_\chi^2/H^2 \sim O(\epsilon)$ and also check that an inflationary solution can be obtained.  The explicit expression for $m^2_\chi$ is given as  
\begin{equation}\label{mChi}
m_{\chi}^2 = \frac{8f_\chi^2fH^2(3-\epsilon)+12f_{\chi\chi}f^2H^2(2-\epsilon)-2f^2V_{\chi\chi}+f^2G_{\phi\phi,\chi\chi}\dot{\phi}^2}{2f(3f_\chi^2+f)}.
\end{equation}
As such, requiring $m_\chi^2/H^2 \sim O(\epsilon)$ can either be achieved by requiring each term individually to be small or by a cancellation amongst the terms.  
Let us consider a simple example.  Taking our background trajectory to be along $\chi = 0$ and 
\begin{align}
V = a(\phi)\chi + b(\phi)\chi^2 + V_0(\phi), \quad\quad f = \frac{1}{2}e^{c\chi}, \quad\quad \mbox{and} \quad\quad G_{\phi\phi} = e^{\phi+c\chi},
\end{align}
we find that $fG_{\phi\phi,\chi} = G_{\phi\phi}f_\chi$ and also $f_\chi R = V_\chi$ so long as $a(\phi) = cR/2$.  Turning to $m_{\chi}^2$ we have 
\begin{equation}
m_{\chi}^2 = \frac{4}{2+3c^2}\left\{c^2H^2(6-2\epsilon) -b(\phi)\right\},
\end{equation}
from which we see that we are able to achieve $m_{\chi}^2/H^2\sim O(\epsilon)$ if $b(\phi) = c^2H^2(6-2\epsilon) \pm O(\epsilon)$ or if $c^2\sim O(\epsilon)$ and $b(\phi)\sim O(\epsilon)$.  Note that this second case agrees with the requirement that $f_\chi/\sqrt{f} = c/\sqrt{2} \sim O(\epsilon^{1/2})$ argued above.    

Finally, we are free to choose our potential $V_0(\phi)$ in order to satisfy the slow-roll conditions \eqref{sr1} and \eqref{sr2}, and note that we are also ``helped'' by a factor $\rm{exp}(-\phi)$ coming from the $1/G_{\phi\phi}$.  For simplicity, taking $V_0 = m^2\phi^2$, from \eqref{slowPhiEom} we get
\begin{equation}
\phi = \ln(e^{\phi_i} - 2\sqrt{m^2/3}t),
\end{equation}
where the subscript $i$ denotes the initial value of $\phi$.  Assuming $e^{\phi_i}\gg2\sqrt{m^2/3}t$, i.e. $\alpha t= 2\sqrt{m^2/3}e^{-\phi_i}t\ll1$, we can expand this as 
\begin{equation}
\phi \simeq \phi_i - \alpha t.
\end{equation}
Using this relation we find an expression for the number of e-foldings $N$ as 
\begin{equation}\label{eFold}
N \simeq \sqrt{\frac{m^2}{3}}\phi_it \simeq \sqrt{\frac{m^2}{3}}\frac{\phi_i}{\alpha}(\phi_i-\phi)=\frac{1}{2}e^{\phi_i}\phi_i(\phi_i-\phi),
\end{equation}  
which in turn gives 
\begin{equation}
a\simeq a_0e^{\sqrt{\frac{m^2}{3}}\phi_i t}.
\end{equation}
From \eqref{eFold} we see that it is relatively simple to achieve $N\gg 60$ (e.g. if $\phi_i \sim O(10)$).

More generally, expanding around the trajectory $\chi = 0$ as
\begin{equation}
V = \sum_nV_{(n)}(\phi)\chi^n,\quad\quad f = \sum_nf_{(n)}\chi^n,\quad\quad\mbox{and}\quad\quad G_{\phi\phi} = \sum_nG^{(n)}_{\phi\phi}(\phi)\chi^n,
\end{equation}
we find 
\begin{equation}\label{mchigen}
 \frac{m_\chi^2}{H^2} = \frac{2}{\left(1+\frac{3f_{(1)}^2}{f_{(0)}}\right)}\left\{2\frac{f_{(1)}^2}{f_{(0)}}(3-\epsilon)+6f_{(2)}(2-\epsilon)-\frac{V_{(2)}}{H^2} +\frac{G_{\phi\phi}^{(2)}\dot{\phi}^2}{2H^2}\right\}.
 \end{equation}
As such, the sufficient condition for realising $m_\chi^2/H^2\sim O(\epsilon)$ is that each term in the numerator of \eqref{mchigen} is $O(\epsilon)$.  Explicitly this gives us
\begin{equation}\label{constraints1}
\frac{f_{(1)}^2}{f_{(0)}}\sim O(\epsilon),\quad\quad f_{(2)}\sim O(\epsilon), \quad\quad \frac{f_{(0)}V_{(2)}}{V_{(0)}}\sim O(\epsilon) \quad\quad\mbox{and}\quad\quad\frac{f_{(0)}G_{\phi\phi}^{(2)}}{G_{\phi\phi}^{(0)}}\sim O(1),
\end{equation}
with the last two being obtained as 
\begin{equation}
\frac{V_{(2)}}{H^2}\simeq\frac{6fV_{(2)}}{V} = \frac{6f_{(0)}V_{(2)}}{V_{(0)}}\sim O(\epsilon)\quad\quad\mbox{and}\quad\quad\frac{G_{\phi\phi}^{(2)}\dot{\phi}^2}{2H^2} = \epsilon\frac{2f_{(0)}G_{\phi\phi}^{(2)}}{G_{\phi\phi}^{(0)}} \sim O(\epsilon),
\end{equation}
where we have used the second relation of \eqref{Hdot} and the definition of $\epsilon$ in \eqref{slowRP}.  Note that the first constraint of \eqref{constraints1} coincides with the condition that $\zeta - \tilde{\zeta}$ is of the correct order of magnitude discussed above.  In addition, from the constraint equations $f_{(0)}G_{\phi\phi}^{(1)}=G_{\phi\phi}^{(0)}f_{(1)}$ and $f_{(1)}R = V_{(1)}$, we obtain
\begin{equation}\label{constraints2}
\frac{V_{(1)}}{V_{(0)}}\sim O(\epsilon^{1/2})\quad\quad\mbox{and}\quad\quad\frac{G_{\phi\phi}^{(1)}}{G_{\phi\phi}^{(0)}} \sim O(\epsilon^{1/2}),
\end{equation}
where we have assumed $f_{(0)} = 1/2$.  
  Of course, the constraints \eqref{constraints1} can be relaxed if there is any cancellation of terms within \eqref{mChi}.    

\section{Discussion} \label{conc}

On considering a multi-field model of inflation with non-minimal coupling and a non-flat field space we have confirmed that, unlike in the single-field case, the curvature perturbation as calculated in the Jordan and Einstein frames are not equivalent.  Furthermore, we were able to explicitly show that the non-equivalence is indeed a direct consequence of the isocurvature perturbations inherent to multi-field models.  As such, in the case that an effectively single-field adiabatic limit is reached, equivalence of the two quantities is recovered.  As a by-product of our formulation, we were also able to confirm that the curvature perturbation is conserved for single-field models, even if the field is non-minimally coupled.  

With the help of a two-field example, we saw that one consequence of the non-equivalence of $\zeta$ and $\tilde{\zeta}$ is that the notion of adiabaticity is not conformally invariant.  This leads to the possibility that whilst in one frame the evolution is adiabatic, and thus the curvature perturbation conserved on super-horizon scales, in the other frame this may not be the case.  We further saw that one could relatively easily obtain an inflationary solution where there was a {\it constant} difference between the curvature perturbation as calculated in the two frames.  Assuming that an effectively single-field adiabatic limit is eventually reached, i.e. equivalence is recovered, and further assuming that the curvature perturbation continues to be conserved in the Einstein frame, we find that the interpretation of the evolution of the curvature perturbation is very different in the two frames.  In the Jordan frame there is a phase in which isocurvature perturbations source the curvature perturbation, whilst in the Einstein frame no such sourcing takes place.  See Fig.\ref{example}.  This highlights the fact that, despite being gauge-invariant, $\zeta$($\sim \mathcal{R}_c$ for $k\ll aH$) is not directly observable, and thus we should be wary when making physical interpretations.  The non-equivalence is also important when it comes to introducing Standard Model matter into the system, which is what we will eventually observe.  If introduced with minimal coupling in the Jordan frame, then the interactions between the $N$ scalar fields and this matter induced by the conformal transformation need to be carefully kept track of, as demonstrated in \cite{ND+MS example}.  If done correctly then the equivalence of observational predictions made in the two frames should be recovered.     

In terms of future work, it would be interesting to analyse the perturbation spectrum of \eqref{JAc} to higher orders, in order to determine the signatures of the non-minimal coupling in non-gaussianities of the power spectrum.  In a similar way as with the first-order perturbations, it would be nice to perform this analysis in a covariant way (with respect to the field-space manifold).  A framework for extending the covariant analysis to higher orders in perturbation theory has recently been discussed in \cite{Tanaka} and \cite{Saffin}.  In terms of relating parameters calculated at the end of inflation with those actually measured in CMB observations, it also seems important to study the reheating process in models with multiple fields and non-minimal coupling.    

\acknowledgments

We would like to thank Seoktae Koh, David Wands and Yuki Watanabe for useful discussions.  JW is supported by the Grant-in-Aid for the Global COE Program ``The Next Generation of Physics, Spun from Universality and Emergence" from the Ministry of Education, Culture, Sports, Science and Technology (MEXT) of Japan.  The work of MM was supported by the Yukawa fellowship and 
by Grant-in-Aid for Young Scientists (B) of JSPS Research, 
under Contract No. 24740162.  This work was also supported in part by JSPS Grant-in-Aid for Scientific Research (A) No.~21244033.

\appendix

\appendixpage


\section{Details of Jordan frame analysis}\label{jFeomApp} 

\subsection{Background and perturbed Einstein equations and equations of motion}
In the Jordan frame, the quantities $\rho$, $p$, $\delta\rho$, $\delta q$, $\delta p$ and $p\Pi_T$ are found to be 
\begin{align}\nonumber
\rho&=\frac{1}{2\kappa^2 f}
\Big[
\frac{1}{2}G_{IJ}
\dot{\phi}^I
\dot{\phi}^J
+V-6H\dot{f}
\Big],\\\nonumber
p&=\frac{1}{2\kappa^2 f}
\Big[
\frac{1}{2}G_{IJ}
 \dot{\phi}^I\dot{\phi}^J
-V
+2\ddot{f}
+4H\dot{f}
\Big],\\
%
\delta\rho
&=\frac{1}{2\kappa^2f}
\Big[
G_{IJ}\big(\dot{\phi}^I \delta\dot{\phi}^J
-\dot{\phi}^I\dot{\phi}^J A\big)
+\frac{1}{2}G_{IJ,K}\delta\phi^K\dot{\phi}^I\dot{\phi}^J
+V_K \delta\phi^K
+6\dot{f}\big(-\dot{\mathcal{R}}+2HA\big)
\nonumber\\
&\hspace{1.5cm}-6H \big(\dot{\delta f}+H\delta f\big)
-\frac{2k^2}{a^2} 
\big(\delta f
-\dot{f}a\frac{\sigma_g}{k}
\big)
\Big],
\nonumber\\
\delta q
&=-\frac{1}{2\kappa^2 f}
\Big[
G_{IJ}\dot{\phi}^I\delta\phi^J
+2
\Big(
\dot{\delta f}
-H \delta f
-\dot{f}
 A
\Big)
\Big],
\nonumber\\
\delta p
&=\frac{1}{2\kappa^2 f}
\Big[
G_{IJ}\big(\dot{\phi}^I \delta\dot{\phi}^J
-\dot{\phi}^I\dot{\phi}^JA\big)
+\frac{1}{2}G_{IJ,K}\delta\phi^K\dot{\phi}^I\dot{\phi}^J
-V_K \delta\phi^K
-2\kappa^2 p\delta f
+2\ddot{\delta f}
\nonumber\\
&\hspace{1.5cm}+4H \dot{\delta f}
-2\dot{f}\dot{A}
+4\dot{f}\dot{\mathcal{R}}
-4\big(\ddot{f}+2 H\dot{f}\big)A
+\frac{4k^2}{3a^2}
\big( \delta f
-\dot{f}a\frac{\sigma_g}{k}\big)
\Big],
\nonumber\\\label{expPertQuantJF}
p\Pi_T&=\frac{k^2}{a^2}\frac{\delta f-a\dot{f}\frac{\sigma_g}{k}}
{\kappa^2 f}.
\end{align}
The coefficients $\bm{M}_2$, $\bm{M}_1$ and $\bm{M}_0$ for the perturbed equations of motion \eqref{jFeomForm} are explicitly given as 
\begin{align}\nonumber
\bm{M}_2 &= 1-\left(A_0\dot{\bm{\phi}}+A_1\bm{\nabla^\dagger}f\right)\bm{\nabla}f \\\nonumber
\bm{M}_1 &= 3H-\bm{\nabla}^\dagger f\left(B_0\bm{\nabla} f + B_1\dot{\bm{\phi}}\bm{\nabla\nabla}f + B_2\dot{\bm{\phi}}^\dagger\right)-\dot{\bm{\phi}}\left(B_3\bm{\nabla}f +B_4\dot{\bm{\phi}}\bm{\nabla\nabla}f\right)-B_5\frac{D\dot{\bm{\phi}}}{dt}\bm{\nabla}f \\\nonumber
\bm{M}_0&=\frac{k^2}{a^2} + \bm{\nabla^\dagger\nabla}(V-fR)-\bm{R}(\dot{\bm{\phi}},\dot{\bm{\phi}})\\\nonumber
&\quad -\bm{\nabla}^\dagger f\left(C_0\bm{\nabla}f + \left(C_1\frac{D\dot{\bm{\phi}}}{dt}+C_2\dot{\bm{\phi}}\right)\bm{\nabla\nabla}f + C_3\dot{\bm{\phi}}^\dagger + C_4\frac{D\dot{\bm{\phi}}^\dagger}{dt} + C_5\bm{\nabla\nabla\nabla}f(\dot{\bm{\phi}}, \dot{\bm{\phi}})\right)\\\nonumber
&\quad-\dot{\bm{\phi}}\left(C_6\bm{\nabla}f + \left(C_7\frac{D\dot{\bm{\phi}}}{dt}+C_8\dot{\bm{\phi}}\right)\bm{\nabla\nabla}f + C_9\dot{\bm{\phi}}^\dagger + C_{10}\frac{D\dot{\bm{\phi}}^\dagger}{dt} + C_{11}\bm{\nabla\nabla\nabla}f(\dot{\bm{\phi}}, \dot{\bm{\phi}})\right)\\
&\quad-\frac{D\dot{\bm{\phi}}}{dt}\left(C_{12}\bm{\nabla}f + C_{13}\dot{\bm{\phi}}\bm{\nabla\nabla}f + C_{14}\dot{\bm{\phi}}^\dagger\right),
\end{align}
where we have adopted an index-free notation for tidiness.  To clarify, a dagger indicates the dual of a vector, such that $(\dot{\phi}^\dagger)_I = G_{IJ}\dot{\phi}^J$ or $(\bm{\nabla}^\dagger V)^I = G^{IJ}\bm{\nabla}_J V$, and, for example, $[\dot{\bm{\phi}}(\dot{\bm{\phi}}\bm{\nabla}\bm{\nabla}f)]^I{}_K = \dot{\phi}^I\dot{\phi}^J\nabla_K(\nabla_Jf)$.  We also have
\begin{align}\nonumber
&[\bm{R}(\dot{\bm{\phi}}, \dot{\bm{\phi}})]^I{}_K = R^I{}_{JLK}\dot{\phi}^J\dot{\phi}^L\\\nonumber
\mbox{and}\quad\quad&[\bm{\nabla\nabla\nabla}f(\dot{\bm{\phi}}, \dot{\bm{\phi}})]_K = \dot{\phi}^I\dot{\phi}^J\nabla_I\nabla_J\nabla_K f,
\end{align}
where $R^I{}_{JKL}$ is the Riemann tensor associated with the metric $G_{IJ}$.  Explicit expressions for the coefficients $A_0$ to $C_{14}$ are given as
\begin{align}\nonumber
&A_0 =C_7 = C_{10} = C_{11} = C_{14}= \frac{1}{2}B_4= \frac{1}{2}B_5= \frac{1}{2}C_{13} = \frac{1}{\dot{f}+2fH}\\\nonumber
&A_1 =C_1=C_5= \frac{1}{2}B_1=3C_{12} = -\frac{6H}{\dot{f}+2fH}\\\nonumber
&B_0 = -\frac{2}{f(\dot{f}+2fH)^2}\left\{2H\dot{f}^2+\dot{f}\ddot{f}-4fH\ddot{f}+H^2f\dot{f}+8\dot{H}f\dot{f}+18f^2H^3+4\dot{H}f^2H\right\}\\\nonumber
&B_2=-\frac{1}{f(\dot{f}+2fH)^2}\left\{\dot{f}^2+4f\dot{f}H+4f^2H^2\right\}\\\nonumber
&B_3=-\frac{1}{(\dot{f}+2fH)^2}\left\{4\dot{H}f+2\ddot{f}-\dot{f}H- 10fH^2\right\}\\\nonumber
&C_0 = -\frac{1}{f(\dot{f}+2fH)^2}\Bigg\{6fH(\dot{f}+2fH)\frac{k^2}{a^2}-2H\dot{f}\ddot{f}+8\ddot{f}fH^2-4\dot{f}^2H^2 -8f\dot{f}H^3\\\nonumber
&\hspace{4cm}-48f^2H^4-16\dot{f}\dot{H}fH+6\dot{H}\dot{f}^2-32\dot{H}f^2H^2\Bigg\}\\\nonumber
&C_2 = -\frac{1}{f(\dot{f}+2fH)^2}\bigg\{2\ddot{f}\dot{f}-8f\ddot{f}H+4\dot{f}^2H+2f\dot{f}H^2+36f^2H^3+8f^2\dot{H}H+16\dot{f}\dot{H}f\bigg\}\\\nonumber
&C_3 = -\frac{1}{f(\dot{f}+2fH)^2}\bigg\{\ddot{f}\dot{f}-4f\ddot{f}H+2\dot{f}^2H+4f\dot{f}H^2+24f^2H^3+4f^2\dot{H}H+8\dot{H}f\dot{f}\bigg\} \\\nonumber
&C_4 = \frac{1}{f(\dot{f}+2fH)^2}\left\{\dot{f}^2-2f\dot{f}H-8f^2H^2\right\}\\\nonumber
&C_6 = \frac{1}{(\dot{f}+2fH)^2}\bigg\{\frac{k^2}{a^2}(\dot{f}+2fH)+2\ddot{f}H-2H^2\dot{f}-12H^3f-4\dot{H}\dot{f}-4f\dot{H}H\bigg\}\\\nonumber
&C_8 = -\frac{1}{(\dot{f}+2fH)^2}\left\{2\ddot{f}-\dot{f}H-10fH^2+4\dot{H}f\right\}\\
&C_9=-\frac{1}{(\dot{f}+2fH)^2}\left\{\ddot{f}-\dot{f}H-6fH^2+2\dot{H}f\right\}.
\end{align}

As a check we consider the minimally coupled case, where $f = 1/2$ (taking $\kappa^2 = 1$).  In this case,  the only non-zero contributions are from
\begin{align}\nonumber
C_9 = -\frac{1}{H^2}(\dot{H}-3H^2) \quad\quad\mbox{and}\quad\quad C_{10} = C_{14} = \frac{1}{H}
\end{align}
which gives us the known result \eqref{covDecEomMinCoup} \cite{Tegmark}.

\subsection{Non-conservation of the curvature perturbation}

In order to determine the behaviour of $\dot{\zeta}$ on large scales, we need to determine $\delta p_{\rm nad}$.  As is obtained in \cite{kt}, on taking the longitudinal gauge ($B = H_T = 0$) and substituting the results \eqref{expPertQuantJF} into \eqref{nadPdef}, we obtain
\bea
-2\kappa^2 f\delta p_{\rm nad}
&=&
\frac{2V_I \dot{\phi}^I}{3H(\rho+p)}\delta\rho
+2 V_I \delta\phi^I  
+{\cal F}
-\delta{\cal F}
+2\kappa^2 f {\cal S}\delta\rho
\nonumber\\
&=&
\frac{2V_I \dot{\phi}^I}{3H(\rho+p)}\delta\rho_m
+2 V_I \Delta^I  
+{\cal F}
-\delta{\cal F}
+2\kappa^2 f {\cal S}\delta\rho,
\eea
where 
\bea
{\cal F}&=&
 4\ddot{f} A
+2\dot{f} \dot{A}
+10\dot{f}
\big(-\dot{\mathcal{R}}+2HA\big),
\nonumber\\
\delta{\cal F}
&=&
 2\kappa^2 \big(\rho-p\big)\delta f
+2\ddot{\delta f}
+10H\dot{\delta f}
+\frac{10}{3}\frac{k^2}{a^2}\delta f,
\nonumber\\
{\cal S}
&=&\frac{\dot{f}}{2Hf}
\Big(1+\frac{4p}{3(\rho+p)}\Big)
+\frac{1}{6 H\dot{H} f}
\Big(
\dddot{f}
+5H\ddot{f}
\Big),
\eea
and 
\bea\label{sFpCg}
\Delta^I
:=\delta\phi^I
+\frac{\delta q}{\rho+p}
\dot{\phi}{}^I
\eea
denotes the scalar field perturbation
in the comoving gauge.  Our notation here exactly follows that of \cite{kt}, the only difference being that explicit expressions for $\delta\rho$ and $\delta q$ now contain additional terms resulting from the non-canonical kinetic factor $G_{IJ}$.  From \eqref{poisson} we know that the $\delta\rho_m$ term can be ignored on super-horizon scales
and hence
\bea
-2\kappa^2f
\delta p_{\rm nad}
\approx 
2 V_I \Delta^I  
+{\cal F}
-\delta {\cal F}
+2\kappa^2 f {\cal S}\delta\rho.
\label{pnad}
\eea

In its current form, as pointed out in \cite{kt}, \eqref{pnad} nicely highlights the contribution to non-conservation of the curvature perturbation due to the non-minimal coupling, with the last three terms vanishing in the minimally coupled case.  However, we would now like to try and re-express \eqref{pnad} in a way that also allows for an intuitive interpretation with regard to the distinction between single and multi-field models.  By introducing the gauge-invariant variables \eqref{JHdefs} we are able to express $\delta p_{\rm{nad}}$ in the form of  \eqref{dPnadFormJF}, with the coefficients given explicitly as
\begin{align}\nonumber
\mathcal{N}_{JK} &= \frac{1}{2\kappa^2f}\Bigg\{\frac{2V_K\big((G_{IJ}+2f_{IJ})\dot{\phi}^I-2H f_J\big)}{2\kappa^2f(\rho+p)}+
(5+3\mathcal{S})H \frac{2f_K}{\dot{f}}f_{IJ}\dot{\phi}^I-\frac{3f_K}{f}f_{IJ}\dot{\phi}^I\\\nonumber&\hspace{2cm}+\frac{2f}{3\dot{f}}\frac{d}{dt}\left(\frac{3f_K}{f}f_{IJ}\dot{\phi}^I\right)-\frac{f_K}{f}G_{IJ}\dot{\phi}^I\Bigg\},\\\nonumber
\mathcal{P}_{JK}&= \frac{1}{2\kappa^2f}\Bigg\{\frac{2f_K}{\dot{f}}f_{IJ}\dot{\phi}^I\Bigg\},\\\nonumber
\mathcal{Q}_{JK}&=\frac{1}{2\kappa^2f}\Bigg\{\frac{3f_Jf_K}{f}-\frac{4V_Jf_K}{2\kappa^2f(\rho+p)}-(5+3\mathcal{S})\frac{2f_Jf_K}{\dot{f}}-\frac{2f}{3\dot{f}}\frac{d}{dt}\left(\frac{3f_Jf_K}{f}\right)\Bigg\},\\
\mathcal{T}_{JK}&=-\frac{1}{2\kappa^2f}\Bigg\{\frac{2f_Jf_K}{\dot{f}}\Bigg\}.
\end{align}

\subsection{Two-field curvaton-like example coefficients}

In the two-field example of Sec.\ref{2fe}, the coefficients $c_\phi$, $m_\phi^2$, $\gamma$, $\lambda$, $c_\chi$ and $m_\chi^2$ appearing in \eqref{reddChiEom} and \eqref{reddPhiEom} are given as
{\allowdisplaybreaks \begin{align}\nonumber
c_\phi &= \frac{G_{\phi\phi,\phi}\dot{\phi}}{G_{\phi\phi}}\\\nonumber
m_\phi^2 &= \frac{G_{\phi\phi}\left(4f\dot{H}^2-H\dot{\phi}(3HG_{\phi\phi}\dot{\phi}+2V_\phi)\right)-fH^2\left(2V_{\phi\phi}+G_{\phi\phi,\phi\phi}\dot{\phi}^2-\frac{G_{\phi\phi,\phi}}{G_{\phi\phi}}(2V_\phi+G_{\phi\phi,\phi}\dot{\phi}^2)\right)}{2fG_{\phi\phi}H^2}\\\nonumber
\gamma&=\frac{f_\chi}{4f^2G_{\phi\phi}(3f_\chi^2+fG_{\chi\chi})H^2}\Bigg\{3f_\chi^2G_{\phi\phi}^2\dot{\phi}^3 + f\left(G_{\phi\phi}^2G_{\chi\chi}\dot{\phi}^3+6f_\chi^2H(G_{\phi\phi,\phi}\dot{\phi}^2+2G_{\phi\phi}\ddot{\phi})\right)\\\nonumber&\hspace{5cm}+2f^2H\left(G_{\phi\phi,\phi}G_{\chi\chi}\dot{\phi}^2 + G_{\phi\phi}(2G_{\chi\chi}\ddot{\phi}-G_{\chi\chi,\phi}\dot{\phi}^2)\right)\Bigg\}\\\nonumber
\lambda &= -\frac{1}{4f^2G_{\phi\phi}(3f_\chi^2+fG_{\chi\chi})H}\Bigg\{4f^3G_{\phi\phi,\phi\chi}G_{\chi\chi}H\dot{\phi}^2-f_\chi^3G_{\phi\phi}^2\dot{\phi}^3\\\nonumber& \hspace{2cm}+ f^2f_\chi\Big(6H\dot{\phi}^2(2f_\chi G_{\phi\phi,\phi\chi}-G_{\phi\phi,\phi}G_{\chi\chi})-24f_{\chi\chi}G_{\phi\phi}H^2\dot{\phi}\\\nonumber&\hspace{2cm}+G_{\phi\phi}(2V_{\chi\chi}\dot{\phi}-G_{\phi\phi,\chi\chi}\dot{\phi}^3-4G_{\chi\chi}H\ddot{\phi})\Big)\\\nonumber&\hspace{2cm}-ff_\chi\left(G_{\phi\phi}^2(G_{\chi\chi}-3f_{\chi\chi})\dot{\phi}^3+6f_\chi^2H\left(3G_{\phi\phi,\phi}\dot{\phi}^2+2G_{\phi\phi}(2H\dot{\phi}+\ddot{\phi})\right)\right)\Bigg\}\\\nonumber
c_\chi& = \frac{f G_{\chi \chi,\phi} \dot{\phi}}{ 3 f_\chi^2+f G_{\chi \chi}}\\
m^2_\chi& = -\frac{2f_\chi^2G_{\phi\phi}\dot{\phi}^2+f(3f_{\chi\chi}G_{\phi\phi}\dot{\phi}^2-24f_\chi^2H^2)-f^2(24f_{\chi\chi}H^2-2V_{\chi\chi}+G_{\phi\phi,\chi\chi}\dot{\phi}^2)}{2f(3f_\chi^2+fG_{\chi\chi})}
\end{align} }
and $m^2_{\phi tot}$, $m^2_{\chi tot}$, $\alpha$ and $\beta$ appearing in \eqref{quantchi} and \eqref{quantphi} are given as 
\begin{align}\nonumber
m_{\chi \rm{tot}}^2 &= m_\chi^2 + \frac{1}{a^2}\left(aH\frac{G_{\chi\chi}^\prime}{G_{\chi\chi}+3f_\chi^2/f} - \frac{1}{4}\left(\frac{G_{\chi\chi}^\prime}{G_{\chi\chi}+3f_\chi^2/f}\right)^2 + \frac{1}{2}\frac{G_{\chi\chi}^{\prime\prime}}{G_{\chi\chi}+3f_\chi^2/f}\right)\\\nonumber
m_{\phi \rm{tot}}^2 &=m_\phi^2 + \frac{1}{a^2}\left(aH\frac{G_{\phi\phi}^\prime}{G_{\phi\phi}} - \frac{1}{4}\left(\frac{G_{\phi\phi}^\prime}{G_{\phi\phi}}\right)^2 + \frac{1}{2}\frac{G_{\phi\phi}^{\prime\prime}}{G_{\phi\phi}}\right)\\\nonumber
\alpha&=a\gamma\sqrt{\frac{G_{\phi\phi}}{G_{\chi\chi}+3f_\chi^2/f}}\\
\beta&= \left\{a^2\lambda-a\gamma\left(aH+\frac{1}{2}\frac{G_{\chi\chi}^\prime}{G_{\chi\chi}+3f_\chi^2/f}\right)\right\}\sqrt{\frac{G_{\phi\phi}}{G_{\chi\chi}+3f_\chi^2/f}},
\end{align}
where here a prime denotes differentiation with respect to conformal time.

\section{Details of Einstein frame analysis}\label{EFApp}

In the Einstein frame, the quantities $\tilde{\rho}$, $\tilde{p}$, $\delta\tilde{\rho}$, $\delta \tilde{q}$, $\delta \tilde{p}$ and $\tilde{p}\tilde{\Pi}_T$ are found to be 
%
%
\begin{align}\nonumber
\tilde{\rho} &= \frac{1}{2}S_{IJ}\phi^{\prime I}\phi^{\prime J} + \frac{V}{(2\kappa^2f)^2},\\\nonumber
\tilde{p} &= \frac{1}{2}S_{IJ}\phi^{\prime I}\phi^{\prime J} - \frac{V}{(2\kappa^2f)^2},\\\nonumber
%
%
\delta \tilde{\rho} &= \frac{1}{2}\delta S_{IJ}\phi^{\prime I}\phi^{\prime J}+ S_{IJ}\delta\phi^{\prime I}\phi^{\prime J}
 + \frac{2\tilde{A}\tilde{H}^\prime}{\kappa^2}- \frac{2V\delta f}{(2\kappa^2)^2f^3}+\frac{\delta V}{(2\kappa^2f)^2},\\\nonumber
\delta \tilde{p} &= \frac{1}{2}\delta S_{IJ} \phi^{\prime I}\phi^{\prime J} - \tilde{A} S_{IJ}\phi^{\prime I}\phi^{\prime J} + S_{IJ}\delta\phi^{\prime I}\phi^{\prime J} + \frac{2V\delta f}{(2\kappa^2)^2f^3}-\frac{\delta V}{(2\kappa^2f)^2},\\\nonumber
\delta \tilde{q} &= - \left(S_{IJ}\phi^{\prime I}\delta\phi^J\right),\\\label{EEdE}
\tilde{p}\tilde{\Pi}_T&=0,
\end{align}
where here a prime denotes differentiation with respect to $\tilde{t}$.  

\end{document}